\documentclass{article}
\usepackage[utf8]{inputenc}
\usepackage[margin=1in]{geometry}
\usepackage[nopatch=footnote]{microtype}
\usepackage{subcaption}
\usepackage{graphicx}
\usepackage{amsmath}
\usepackage[hyphens]{url}
\usepackage{hyperref}
\hypersetup{colorlinks=false,breaklinks=true,hidelinks}
\newcommand{\suppfigref}[1]{Supplementary Fig.~\ref*{#1}}

\newcommand{\suppfigtworef}[2]{Supplementary Figs.~\ref*{#1} and~\ref*{#2}}
\newcommand{\supptabref}[1]{Supplementary Table~\ref*{#1}}
\newcommand{\suppnoteref}[1]{Supplementary Note~\ref*{#1}}
\usepackage[natbibapa]{apacite}
\usepackage{authblk}
\usepackage{booktabs,multirow,longtable,makecell,tabularx}
\usepackage{siunitx}
\usepackage{threeparttable}
\usepackage{placeins}
\graphicspath{{figures/}}
\usepackage{setspace}
\title{Event-Centric Earthquake Monitoring with a Multi-Station Transformer}
\author[1,2]{Haoyu Wang\thanks{haoyu.wang@berkeley.edu}}
\author[1,2]{Weiqiang Zhu}
\affil[1]{Department of Earth \& Planetary Science, University of California, Berkeley}
\affil[2]{Berkeley Seismology Laboratory, University of California, Berkeley}

\date{}

\begin{document}

\maketitle

\begin{abstract}
Earthquake catalogs underpin studies such as fault structure, earthquake processes and seismic hazard, and their value depends on being both complete and reliable. Machine learning has improved individual tasks of earthquake monitoring, but processing stations independently discards the consistency of arrivals across the network, and running tasks in sequence fragments the information describing each earthquake. We present EQNet2, a geometry-aware multi-station Transformer that jointly predicts phase arrivals, first-motion polarities, source location and magnitude, and the association of arrivals into events. On a Northern California test dataset and on continuous waveforms from Ridgecrest and the San Francisco Bay Area, EQNet2 recovers more weak arrivals, produces fewer false detections, and associates more picks into located events. Its learned attention links information across stations along time--distance patterns consistent with seismic moveout. As seismic networks grow denser, exploiting multi-station coherence and event-centric prediction offers a path to more complete and reliable catalogs.
\end{abstract}

\section{Introduction}

Earthquake monitoring converts continuous waveforms recorded by a seismic network into catalogs that report the timing, location, magnitude, and source mechanism of individual earthquakes. These catalogs support research across seismology, from earthquake source processes and fault structure to seismic hazard. For example, dense and precisely located seismicity images fault geometry at depth \citep{ross2019hierarchical,tan2021machine,poggiali2025fault}. The distribution and migration of seismicity trace fluid movement and aseismic slip \citep{ross2020fault,ross2021evidence,park2020machine,wang2026reactivation,suzuki2025forearc}. Seismicity beneath volcanoes delineates the structures activated during unrest and the pathways of magma transport \citep{tan2025clearer,wilding2023magmatic,lomax2025santorini}. Catalogs with a lower magnitude of completeness also resolve the statistics of earthquake sequences such as early-aftershock migration, precursory seismicity, and b-value variations \citep{peng2009migration,trugman2019pervasive,kwiatek2023months,herrmann2022revealing}.
These results depended on both the quantity and the quality of the detected earthquakes. Missed detections raise the magnitude of completeness \citep{woessner2005assessing,schorlemmer2008probability}, false detections inflate event rates and add spatial scatter, and location errors blur fault structure and may introduce location artifacts \citep{waldhauser2000double,hauksson2005southern,hauksson2012waveform}. Producing high-resolution catalogs that are complete, reliable, and precise therefore depends on continued advances in automated monitoring workflows.

Automated monitoring workflows convert continuous waveforms into event-level information through a sequence of tasks: picking P- and S-phase arrival times at each station, determining first-motion polarities, associating picks that originate from a common source, locating hypocenters, and estimating magnitudes and focal mechanisms. Deep learning has substantially improved each of these tasks \citep{mousavi2022deep}. Deep-learning phase pickers trained on millions of analyst-labeled seismograms now pick arrivals with near-analyst accuracy and detect weaker phases than traditional algorithms \citep{ross2018generalized,zhu_phasenet,zhou2019hybrid,mousavi_eqtransformer_2020,soto2021deepphasepick,saad2023eqcct,chen2024eqcct}. Convolutional classifiers determine first-motion polarities that previously required manual inspection, enabling focal mechanisms for abundant small events \citep{ross2018p,hara2019p,mousavi2019unsupervised,chakraborty2022polarcap,cheng2023refineda}. Machine-learning associators efficiently group large pick sets produced by deep-learning pickers into candidate events \citep{ross2019phaselink,mcbrearty2019pairwise,zhu_gamma_2022,mcbrearty_genie}. Location models provide fast preliminary hypocenters \citep{mousavi2019location,mcbrearty2022location,zhu2025robust}, and transformer models determine focal mechanisms from more complete seismic information \citep{song2026foconet}. Chaining these tools into pipelines has produced a new generation of high-resolution catalogs from dense and long-duration seismic recordings \citep{zhang2022locflow,zhu2023quakeflow,tan2021machine,wilding2023magmatic,kwiatek2023months,zhang2024landslide,feng2024microseismic,tan2025clearer}.

Despite these advances, two limitations of current automated monitoring workflows remain. The first is that most phase pickers still process each station independently \citep{chen2022cubenet,sun2023phaseno}, whereas an earthquake is recorded by the network as a whole. Weak arrivals that are ambiguous on a single noisy seismogram become identifiable once their timing aligns with the moveout of arrivals at neighboring stations, and local noise that resembles a phase arrival on one trace has no coherent arrivals elsewhere in the network. Single-station pickers cannot use this consistency across the network, so they miss weak arrivals and pass isolated false picks to association. The second limitation is that the workflow is a unidirectional sequence of independently trained modules, so the workflow as a whole is not optimized for catalog quality \citep{zhu_eqnet_2022,si2024plan}. Each stage passes only discrete predictions to the next, without the waveforms or the uncertainties behind them, so errors made early are difficult to correct later and accumulate downstream.

Recent work has approached these two limitations. Multi-station pickers such as CubeNet \citep{chen2022cubenet}, EdgePhase \citep{feng2022edgephase}, and PhaseNO \citep{sun2023phaseno} take waveforms from many stations as joint input, so that the pick at one station is informed by the waveforms at the others. Multi-task models predict several attributes of the same event from a shared representation. PhaseNet+ \citep{zhu2025towards} adds first-motion polarity and origin-time estimation to phase picking, and an attention-based model picks arrivals and determines polarity simultaneously \citep{zhang2023simultaneous}. The two directions have begun to converge. EQNet \citep{zhu_eqnet_2022} performs picking and association jointly from multi-station input, and PLAN \citep{si2024plan} adds location. These studies show that automated monitoring improves both when information is shared across stations and when related tasks are predicted jointly. Extending this combination to the entire workflow could overcome both limitations at once, recovering weak arrivals and rejecting isolated noise through network coherence while keeping every measurement tied to the same earthquake.

In this study, we combine network-wide input with event-centric prediction in a single model, EQNet2, a multi-station Transformer that learns features from multiple stations and organizes its predictions around individual earthquakes (Figure~\ref{fig:workflow}). EQNet2 jointly predicts phase arrivals, first-motion polarities, source location and magnitude, and the association of arrivals into events from a shared representation of multi-station waveforms and relative station geometry. A learnable spatiotemporal relative position bias incorporates station geometry into the attention mechanism, allowing the model to accommodate varying station configurations. The model is trained on event-organized waveforms from the California Earthquake Event Dataset \citep{zhu2025california} and evaluated on a Northern California test dataset and on continuous waveforms from the 2019 Ridgecrest sequence and the 2025 Berkeley earthquake. Across these tests, EQNet2 recovers more weak arrivals, produces fewer false detections, and retains a larger fraction of its picks in located events.

\section{Methods}
\label{sec:methods}

\subsection{Event-centric monitoring across the station network}
\begin{figure}[htbp]
    \centering
    \includegraphics[width=\linewidth]{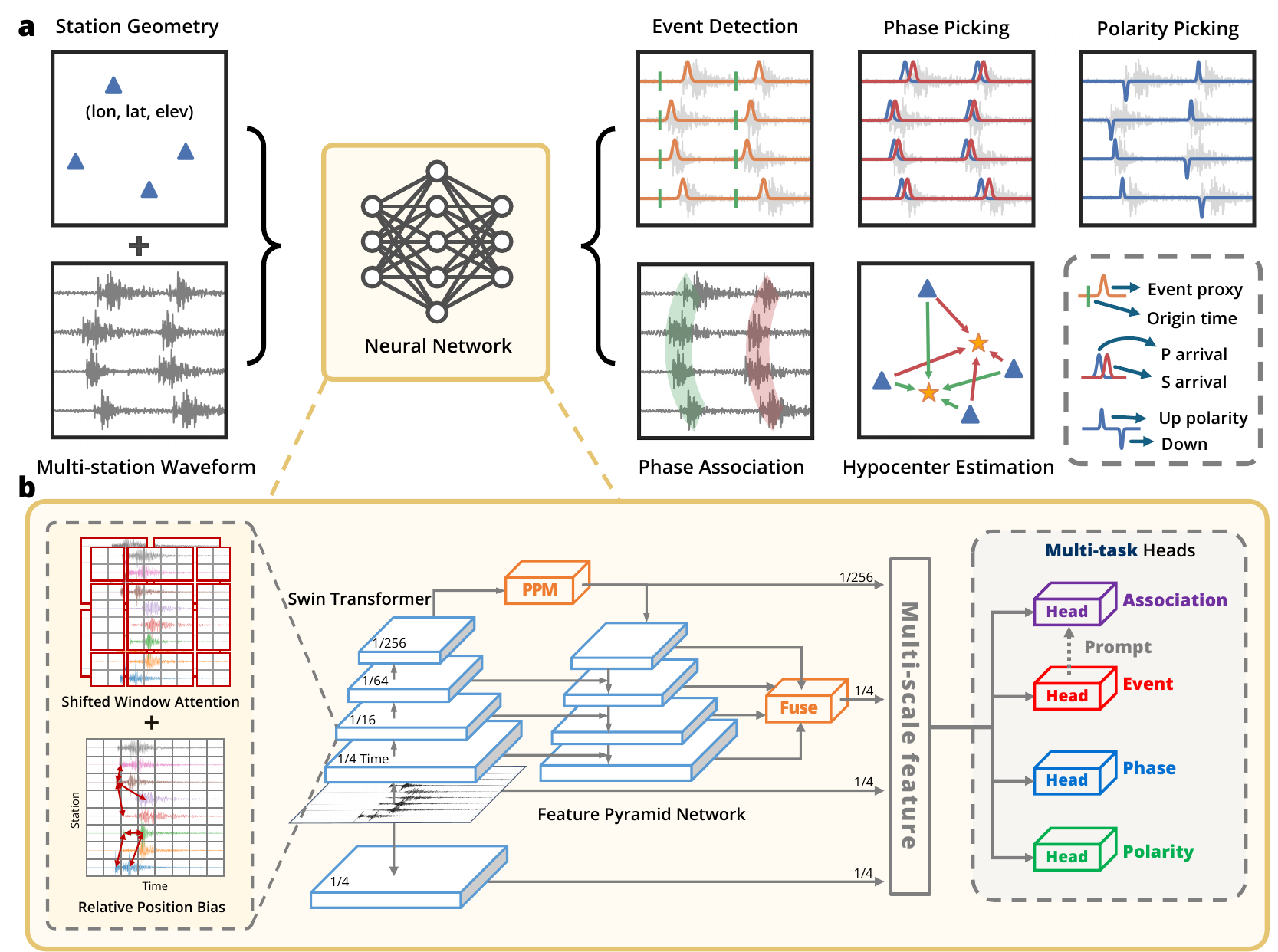}
    \caption{\textbf{Event-centric formulation and geometry-aware architecture of EQNet2.} \textbf{a,} Multi-station waveforms and relative station geometry are encoded into a shared representation from which phase arrivals, first-motion polarities, event-arrival proxies, source-related estimates, and network-wide association probability masks are jointly predicted for the same candidate events. \textbf{b,} Detailed structure of EQNet2. Multi-station waveforms are processed by a hierarchical Swin Transformer with spatiotemporal relative position bias and fused through a feature pyramid network before task-specific decoding. Predicted event-arrival proxies serve as prompts for event-specific association. PPM, pyramid pooling module \citep{zhao2017pyramid}. Detailed architectures of the prediction heads are shown in \suppfigref{fig:heads_arch}.}
    \label{fig:workflow}
\end{figure}

EQNet2 performs the interconnected tasks of earthquake monitoring in one model that processes the whole network at once. It takes multi-station waveforms and station coordinates as input and jointly produces phase arrivals, first-motion polarities, station-wise event-arrival proxies, source-related estimates (travel time, S--P differential time, epicentral distance, backazimuth, depth, and magnitude), and network-wide association probabilities (Figure~\ref{fig:workflow}a). Rather than generating these outputs in separately optimized modules, EQNet2 decodes them all from one representation shared across stations and tasks, so each output refers to the same candidate events. Following PhaseNet+ \citep{zhu2025towards}, the event-arrival proxy is defined as the temporal midpoint between the P- and S-phase arrivals. The proxy gives each trace a reference time at which the source-related estimates are extracted. Each predicted proxy further serves as the prompt for association, linking the arrivals at individual stations to candidate events.

\begin{figure}[htbp]
    \centering
    \includegraphics[width=\linewidth]{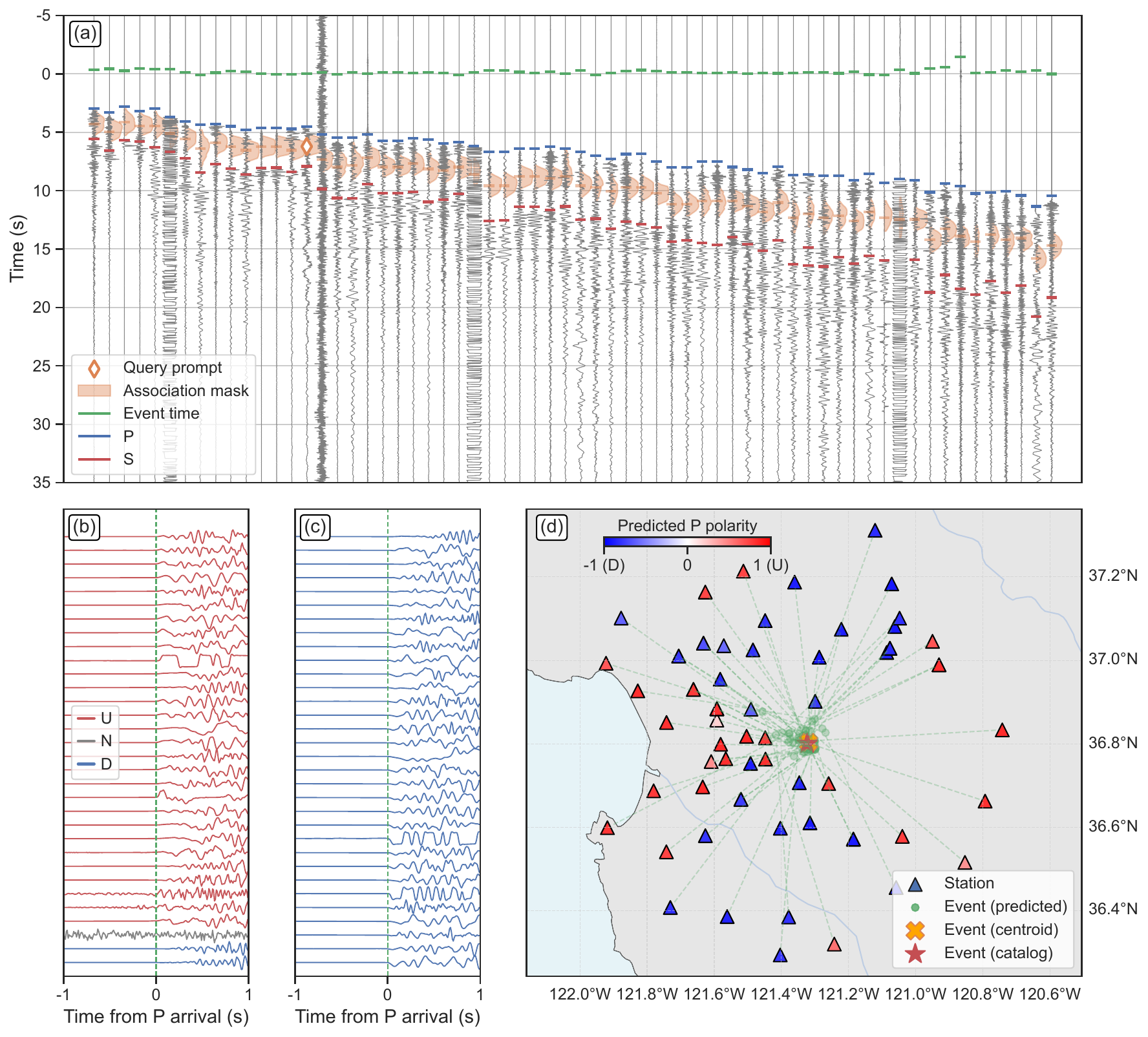}
    \caption{\textbf{Example of event-level outputs from EQNet2.} \textbf{a,} Vertical-component waveforms sorted by source--station distance. Blue and red ticks mark predicted P- and S-phase arrivals, respectively; green ticks mark station-wise origin-time estimates obtained from the predicted event-arrival proxies and travel times. The orange diamond denotes the query event-arrival proxy, and orange shading shows the corresponding association probabilities. \textbf{b,c,} Vertical-component waveforms aligned on the predicted P arrivals and colored by predicted first-motion polarities. Polarities are categorized as Up (U), Unknown (N), or Down (D) using a score threshold of 0.3. \textbf{d,} Seismic stations, station-wise epicentral estimates, their centroid, and the reference epicenter. Dashed lines indicate predicted backazimuths, and station colors indicate predicted first-motion polarities.}
    \label{fig:event_detection}
\end{figure}

Figure~\ref{fig:event_detection} illustrates the joint outputs for event nc73866925 from the NCEDC 2023 test dataset. At each station, EQNet2 predicts P- and S-phase arrivals, a first-motion polarity, and source-related estimates. The arrivals follow a coherent moveout across the network, and the station-wise origin-time estimates cluster around a common value. A single event-arrival prompt query groups these station-level predictions into one candidate event through the association probability mask. The analyses in Section~\ref{sec:results} further quantify the accuracy of each prediction task and its effect on catalog construction.

\subsection{Geometry-aware multi-station representation}
EQNet2 pairs a shared backbone, which extracts features from all stations jointly, with parallel task-specific heads that decode these features into predictions (Figure~\ref{fig:workflow}b).
The backbone is based on the shifted-window (Swin) Transformer \citep{liu_swin_2021,liu_swin2_2022}. It processes multi-station waveform tokens (short waveform segments) with windowed self-attention to balance modeling capacity and computational cost, where each attention window spans a range of time steps and a subset of stations (\suppnoteref{sec:supp_model}). The multiscale features from the Swin backbone are fused by a feature pyramid network (FPN) module \citep{lin2017feature,xiao_upernet}, producing representations suitable for tasks that require different temporal resolutions.

A key component of the backbone is the geometry-aware relative position bias (RPB) in each attention module. The relative arrival times of seismic phases at different stations depend on the geometry of the network relative to the source, whereas a standard attention mechanism \citep{vaswani_transformer_2017} carries no geometric information. EQNet2 therefore adds a learnable bias term to the attention scores so that the exchange of information between waveform tokens can depend on their relative timing and station geometry.
We incorporate the RPB term $B$ into the attention computation:

\begin{equation}
    \operatorname{Attention}(Q,K,V)
    =
    \operatorname{SoftMax}\!\left(\cos(Q,K)/\tau + B\right)V,
\end{equation}

where $Q$, $K$, and $V$ are the query, key, and value feature matrices in the attention module, respectively, and $\tau$ is a learnable temperature parameter that scales the cosine similarity between query and key features.
The bias term $B$ is generated from the relative spatiotemporal coordinates between each pair of waveform tokens:

\begin{equation}
    B = \mathcal{G}(\Delta t,\,R,\,\theta,\,\Delta z),
\end{equation}
where $\Delta t$ is the time difference between two waveform tokens; $R$ is the great-circle distance between the corresponding stations; $\theta$ is the inter-station azimuth, represented by its sine and cosine; and $\Delta z$ is the relative elevation. The function $\mathcal{G}$ is a two-layer multilayer perceptron (MLP). Each Transformer block has an independent $\mathcal{G}$, allowing different layers to learn distinct spatiotemporal interaction preferences. This design lets EQNet2 relate waveforms across stations under irregular network geometry, without a fixed station layout or ordering.

\subsection{Event-centric prediction heads}

Task-specific heads decode the shared multi-station representation into station-level and event-level outputs (\suppfigref{fig:heads_arch}).
The Phase Head outputs per-sample probabilities for noise, P-phase arrival, and S-phase arrival classes, trained as a segmentation task with Gaussian targets centered at manual picks \citep{zhu_phasenet}.
The Polarity Head outputs per-sample probabilities for upward, unknown, and downward first-motion categories, following PhaseNet+ \citep{zhu2025towards}. Because the three polarity classes are ordered (Down, Unknown, Up), we formulate polarity prediction as an ordinal regression task \citep{mccullagh1980regression}.

The Event Head jointly performs event detection and source-related regression tasks. Its detection branch outputs a probability sequence for the event-arrival proxy, which serves as a reference time anchor for extracting phase arrivals and source-related estimates at each station. Separate regression branches predict travel time, S--P differential time, epicentral distance, backazimuth, depth, and magnitude. The regression branches additionally receive raw amplitude information because absolute and relative amplitudes constrain source location and magnitude.

The Association Head groups arrivals into events directly in the same feature space used for detection and picking. Its design is inspired by prompt-based segmentation in the Segment Anything Model (SAM) \citep{kirillov2023segment,ravi2024sam}. Unlike the user-provided spatial prompts in SAM, the EQNet2 prompt is a model-predicted event-arrival proxy. Each candidate proxy is encoded as a query token that attends to backbone features across all stations (\suppfigref{fig:heads_arch}), so the head has access to both waveform features and relative station geometry. Its station positional encoding follows the same relative-geometry design as the backbone RPB, maintaining geometric consistency between the two modules. The output is a network-wide association mask that gives the probability that each detected arrival belongs to the same event as the query (Figure~\ref{fig:event_detection}). During inference, association masks are converted into candidate events by the graph procedure described in \suppnoteref{sec:supp_model}.

\subsection{Training data and optimization}
\label{sec:model_training}

Training a multi-station, multi-task model requires datasets with comprehensive annotations for all relevant tasks linked at the event level.
We use the California Earthquake Event Dataset (CEED) \citep{zhu2025california}, which organizes its records by event, each containing the associated station waveforms, phase picks, polarities, and hypocenter parameters. The Northern California Earthquake Data Center (NCEDC) subset of CEED through 2022 is used for training, and the NCEDC data from 2023 are held out for evaluation. The training subset contains approximately 325,000 events and 1.1M three-component waveform records. Each waveform record spans 120~s at 100~Hz. During training, we randomly select 16 waveform traces per event and crop each trace to a 40.96-s window with a random time shift.
We also retain waveforms without manual labels when at least one trace from the same event contains labels. Such traces occur because not every station recording associated with an event has complete manual annotations, especially for weak phases or low-quality components. Retaining them allows the model to learn from the available waveform context while the supervised losses are applied only to annotated targets.

The model is trained with a composite multi-task loss that combines phase picking, event-arrival-proxy detection, polarity prediction, association, and source-related regression objectives; full loss definitions and optimization settings are provided in \suppnoteref{sec:supp_losses}.
To improve generalization and robustness to complex seismic signals, we employ two data augmentation strategies during training: temporal event stacking and spatial event concatenation. Temporal event stacking superimposes waveform records from different events with randomized time offsets, simulating overlapping arrivals from closely spaced events in time. Spatial event concatenation combines stations from different events within the same training window, simulating cases in which different parts of the network record different nearby sources or unrelated events. These augmentations expose the model to temporally and spatially overlapping events, encouraging it to rely on network coherence when separating event-related signals from unrelated arrivals.

\section{Results}
\label{sec:results}

\subsection{Network information improves the picking of weak arrivals and the estimation of source parameters}
We evaluated EQNet2 on fixed-length waveform windows from the NCEDC 2023 test dataset for phase picking, event detection, first-motion polarity, source-related estimation, and association. We compared EQNet2 with PhaseNet \citep{zhu_phasenet}, PhaseNet+ \citep{zhu2025towards}, and PhaseNO \citep{sun2023phaseno}, as well as a single-station variant that retains the same task formulation but removes cross-station information. Because catalog annotations do not exhaustively label every arrival in each waveform window, a prediction outside the annotated event interval may correspond to an unlabeled neighboring event or an incomplete manual annotation rather than a true false positive. Phase-picking metrics are therefore evaluated only within an event-centered interval extending from 5~s before the catalog P arrival to three times the catalog S--P differential time after the P arrival.

\begin{figure}[htbp]
    \centering
    \includegraphics[width=\linewidth]{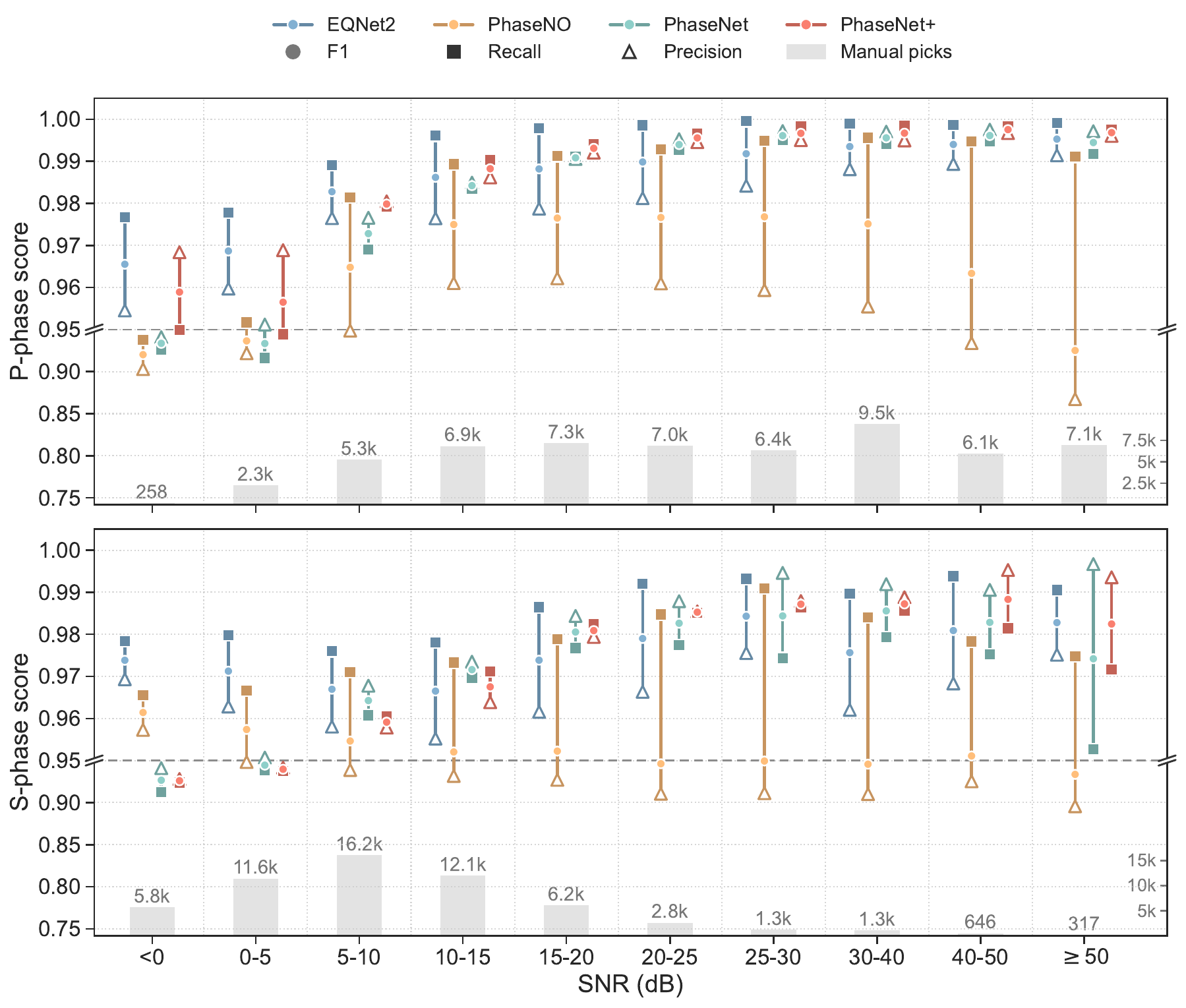}
    \caption{\textbf{Phase-picking performance across signal-to-noise-ratio groups.} Precision (triangles), recall (squares), and F1 score (circles) of P-phase (top) and S-phase (bottom) picking on the NCEDC 2023 dataset. The vertical axis is compressed below 0.95 (dashed line). Gray bars show the number of manual picks in each group (right axis).}
    \label{fig:metrics_vs_snr_f1}
\end{figure}

For phase picking, EQNet2 achieves P-phase performance comparable to the leading baselines and the highest overall S-phase performance (\suppfigref{fig:pr_curve}). EQNet2 also outperforms its single-station variant and produces the smallest standard deviations in the P- and S-phase picking residuals, while maintaining mean residuals comparable to those of the baselines (\suppfigref{fig:metrics_residuals}; \supptabref{tab:phase_picking_metric}). Together, these results indicate that network information improves both recovery and timing precision. We stratified performance by SNR (Figure~\ref{fig:metrics_vs_snr_f1}) and other parameters (\suppnoteref{sec:metrics_vs_params}). EQNet2 achieves the highest recall for both phases in every SNR group, with the largest gains at low SNR. EQNet2 also achieves the highest F1 scores below 10 dB and comparable values in high SNR groups. Measured precision is lower for the multi-station models, partly because the selected evaluation time window can still include uncataloged events in dense scenarios (\suppfigref{fig:compare_test_dataset}).

Beyond phase arrivals, all compared models that include event detection achieve high detection rates, whereas the clearest differences appear in the source-related estimates (\suppfigref{fig:pr_curve}; \supptabref{tab:event_detection_metric}). In particular, the standard deviation of the backazimuth residual decreases from $44.86^\circ$ for the single-station variant to $7.43^\circ$ for multi-station EQNet2 (\suppfigref{fig:metrics_residuals}), consistent with backazimuth being constrained primarily by network geometry rather than any single waveform.

For first-motion polarity, we categorized the predictions of EQNet2 and PhaseNet+ using a threshold of 0.3 (Up $>0.3$, Down $<-0.3$, Unknown otherwise). EQNet2 improves classification of the Up and Down classes relative to PhaseNet+ and the CNN model of \citet{ross2018p} (\suppfigref{fig:polarity_confusion_matrix}). Performance is lower for the ``Unknown'' class. Qualitative inspection suggests that this class is heterogeneous and includes many high-SNR waveforms with visually discernible polarities \citep{zhu2025towards}. Both EQNet2 and PhaseNet+ tend to assign definitive labels (Up or Down) to a subset of these unknown samples, which lowers the Unknown-class accuracy without necessarily indicating an incorrect polarity estimate.

We also evaluated the prompt-based association module against spatiotemporal clustering of the predicted source-related estimates. For each test window, the association ratio measures the fraction of picked arrivals that are assigned to a predicted event group. When $k$ of $n$ picks in a window remain unassociated, the ratio equals $1 - k/n$, and the values therefore fall on discrete curves (\suppfigref{fig:association_ratio}). The prompt-based module leaves picks unassociated in fewer windows, and its advantage grows for windows with many picks, where clustering more often fails to associate several arrivals. Predicted event-arrival proxies are thus effective prompts for gathering picks into events, leaving fewer picks unassociated.

To examine missed detections in dense-event scenarios, we inspected a randomly selected sample from the NCEDC 2023 test dataset containing two cataloged events (\suppfigref{fig:compare_test_dataset}). Among the four compared models, EQNet2 recovers the most complete set of P- and S-phase picks, including weak arrivals whose timing is consistent across stations. Its event-time predictions are also more complete and less scattered than those of PhaseNet+. EQNet2 nevertheless misses some event-arrival proxies and produces occasional isolated S-phase picks in this example. Overall, the example is consistent with the aggregate improvements in S-phase and low-SNR picking observed across the benchmark.

\subsection{Continuous monitoring yields a more complete and spatially coherent Ridgecrest catalog}

Fixed-window performance does not by itself establish catalog quality. We therefore test whether these improvements translate into end-to-end monitoring with continuous waveforms. The 2019 Ridgecrest earthquake sequence featured a $M_w\,7.1$ mainshock on July 5, preceded by a $M_w\,6.4$ foreshock on July 4, followed by a vigorous aftershock sequence that activated a complex fault structure at depth \citep{ross2019hierarchical,shelly2020high}. The high event rate and structural complexity make it a challenging test for detection, association, and overall catalog quality. We processed continuous waveform data from July 4 to July 10, 2019, in a $1^\circ \times 1^\circ$ region centered at 117.504$^\circ$W, 35.705$^\circ$N. The data comprise 45 sensors from 16 CI stations, including HH, EH, HN, and BH channels.
We compare the EQNet2 catalog with the Southern California Seismic Network (SCSN) catalog and with two catalogs built from the same data, one from PhaseNet+ picks associated by predicted origin time \citep{zhu2025towards} and one from PhaseNO picks associated with GaMMA \citep{zhu_gamma_2022}. Associated picks were located using ADLoc \citep{zhu2025robust} with the SCSN 1D velocity model \citep{hutton2010earthquake}.

\begin{figure}[htbp]
    \centering
    \includegraphics[width=\linewidth]{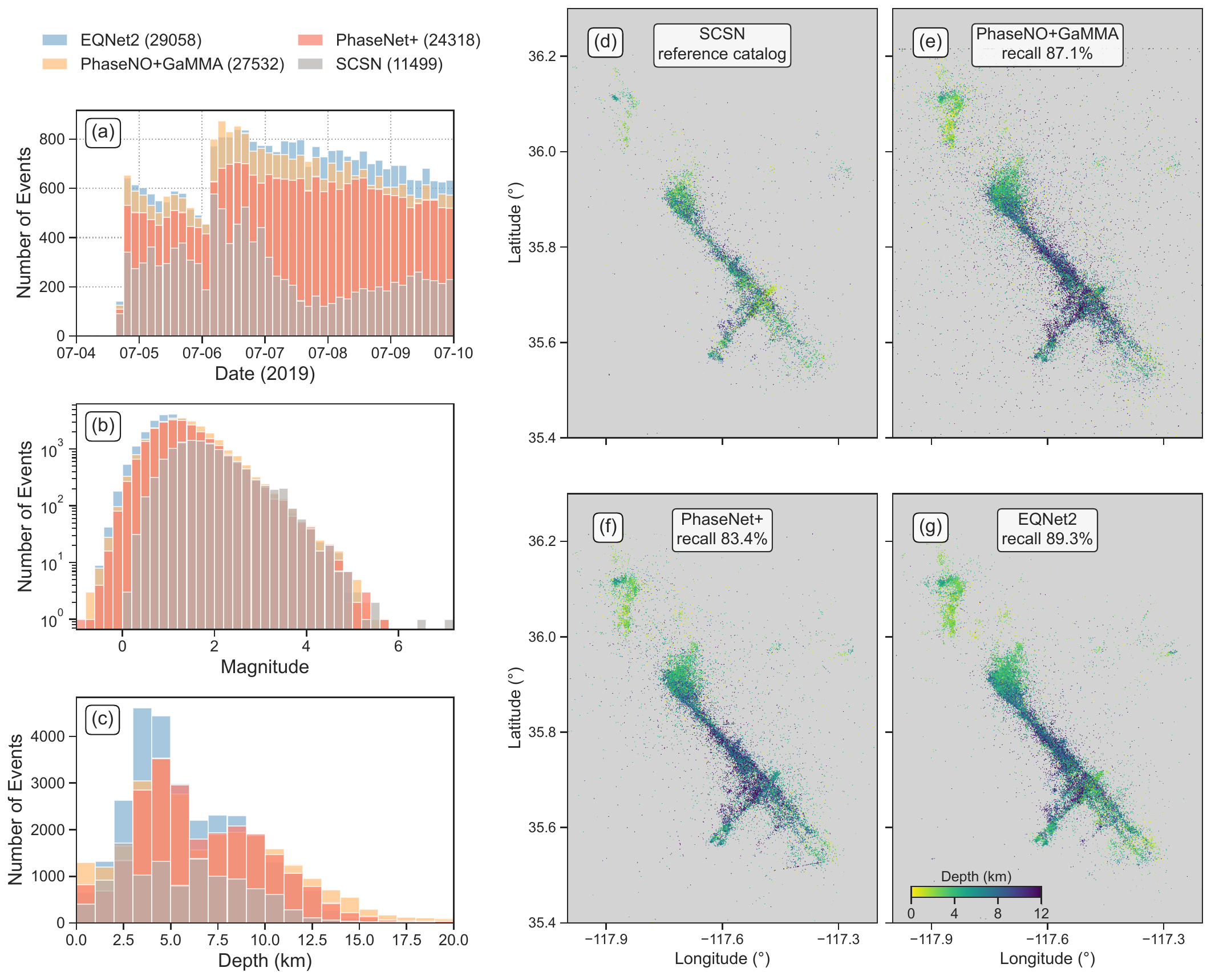}
    \caption{\textbf{Comparison of earthquake catalogs for the 2019 Ridgecrest sequence.} \textbf{a,} Temporal event frequency. \textbf{b,} Magnitude distribution. \textbf{c,} Depth distribution. \textbf{d--g,} Spatial distributions for SCSN, PhaseNO+GaMMA, PhaseNet+, and EQNet2. Event counts are shown in the legend, and labels in panels e--g report event recovery relative to the SCSN catalog. Colors in panels d--g indicate event depth.}
    \label{fig:ridgecrest_combined}
\end{figure}

Figure~\ref{fig:ridgecrest_combined} compares the temporal, magnitude, depth, and spatial distributions across four catalogs. The EQNet2 catalog contains 29,058 events, slightly more than the PhaseNO+GaMMA catalog (27,532), approximately 20\% more than the PhaseNet+ catalog (24,318), and more than twice as many as the SCSN catalog (11,499). Relative to the SCSN catalog, EQNet2 recovers 89.3\% of the events, compared with 87.1\% for PhaseNO+GaMMA and 83.4\% for PhaseNet+. At the pick level, 73.9\% (1.44~M / 1.95~M) of the raw EQNet2 picks are associated into events and used for location, compared with 70.9\% (1.14~M / 1.61~M) for PhaseNO+GaMMA and 59.2\% (0.90~M / 1.52~M) for PhaseNet+. Thus, the larger EQNet2 catalog is supported by both higher recovery of SCSN events and a larger fraction of raw picks contributing to located events. The additional EQNet2 events concentrate at depths of 2--6~km and align with the orthogonal subfault structures identified by \citet{ross2019hierarchical}. The EQNet2 catalog delineates fault structures more clearly with less spatial scatter, particularly toward the southeastern end of the fault system.

Before absolute location, hypocenter estimates obtained directly from EQNet2 already capture the general fault geometry (\suppfigref{fig:ridgecrest_raw_location}). Increasing station-window size from the 16 traces used during training to all 45 traces reduces spatial scatter and yields more concentrated event locations. The accompanying reduction in the number of associated events is consistent with fewer split events when more stations constrain each candidate event.

Using the EQNet2 first-motion polarities, we computed focal mechanisms with SKHASH \citep{skoumal2024skhash}. The resulting compressional-axis (P-axis) azimuths are consistent with the dominant spatial patterns in the SCSN catalog \citep{yang2012computing}, the catalog of \citet{cheng2023refineda}, and the PhaseNet+ catalog (\suppfigref{fig:ridgecrest_fm}). EQNet2 produces more focal-mechanism solutions in the northern and central parts of the sequence, mainly within the interior of the station network where azimuthal coverage is more favorable.

\subsection{Network coherence recovers weak arrivals and suppresses false detections}

The Ridgecrest sequence primarily tests event recovery under dense seismicity. We next examine one day of continuous data around the 2025 Berkeley $M\,4.3$ earthquake, where strong cultural (anthropogenic) noise provides a complementary test of whether network information can simultaneously recover weak event-related arrivals and reject non-earthquake signals. We processed waveforms from 69 stations within a $0.6^\circ \times 0.6^\circ$ region centered on the epicenter.

\begin{figure}[htbp]
    \centering
    \includegraphics[width=\linewidth]{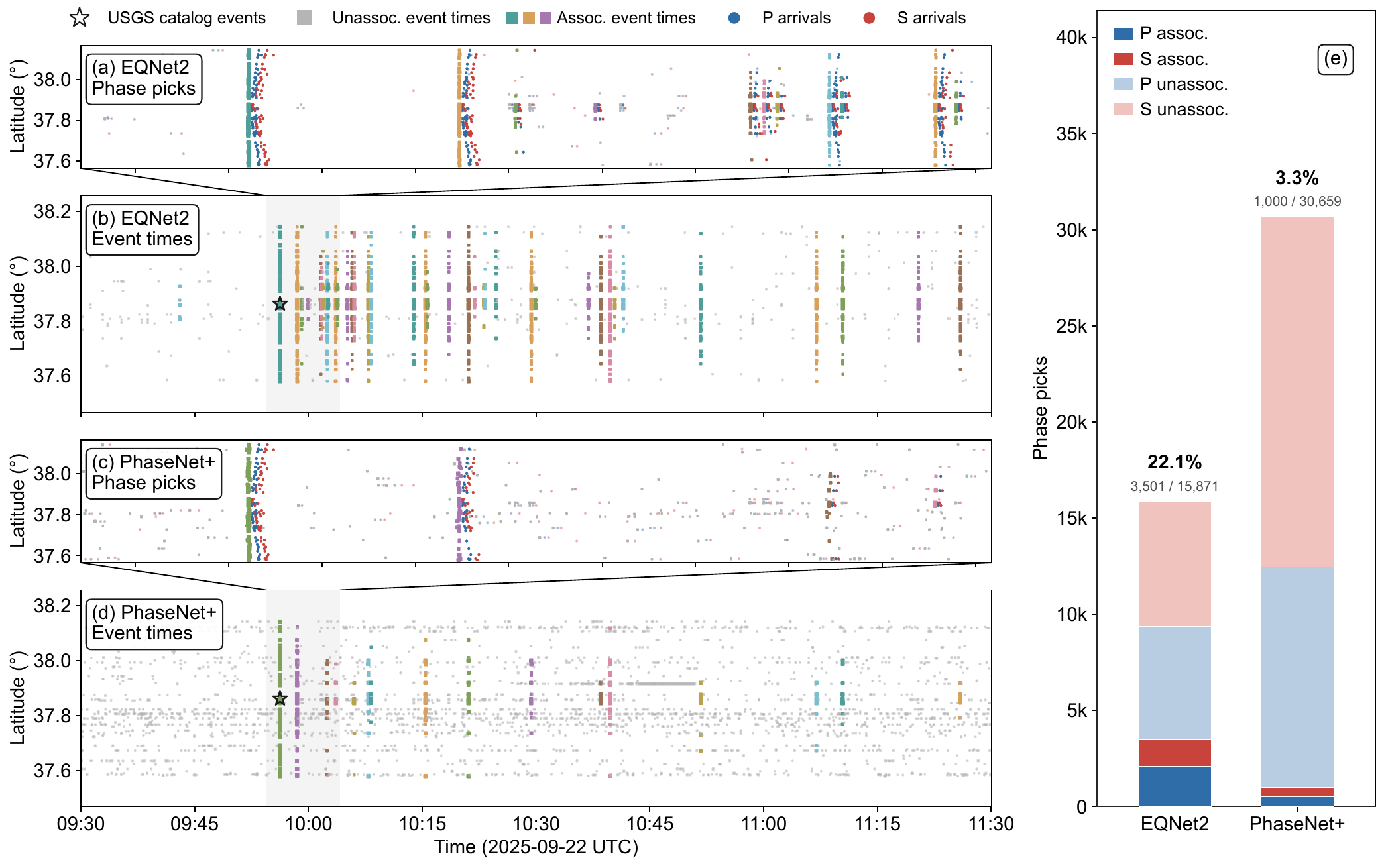}
    \caption{\textbf{Comparison of EQNet2 and PhaseNet+ for the 2025 Berkeley earthquake sequence.} \textbf{a--d,} Predictions in a 2-hour window around the 2025-09-22 Berkeley $M\,4.3$ earthquake (event ID: nc75240492). Each square in panels b and d denotes a predicted event origin time at the corresponding station. Gray squares are unassociated predictions, colored squares indicate predictions associated with the same event, and the black star marks the USGS catalog event. Panels a and c enlarge the shaded interval and additionally show predicted P- and S-phase arrivals. \textbf{e,} Stacked counts of associated and unassociated P- and S-phase picks. Percentages indicate the fraction of raw picks retained after association and location.}
    \label{fig:case_berkeley}
\end{figure}

EQNet2 produces fewer raw P- and S-phase picks than PhaseNet+ (15,871 versus 30,659), but retains a substantially larger fraction after association and absolute location: 3,501 picks for EQNet2 (22.1\%) and 1,000 picks for PhaseNet+ (3.3\%) (Figure~\ref{fig:case_berkeley}e). In the 2-hour window around the mainshock, many PhaseNet+ event predictions remain unassociated and scatter across the network, whereas the EQNet2 predictions concentrate into aftershocks shared across stations. The enlarged panels show the contrast at the phase level, where EQNet2 recovers P- and S-phase arrivals at more stations for each associated event, giving the aftershocks more complete arrival patterns (Figure~\ref{fig:case_berkeley}a--d). Both models recover similar numbers of high-SNR arrivals, whereas the additional located EQNet2 picks are predominantly low-SNR arrivals supported by coincident detections at neighboring stations (\suppnoteref{sec:berkeley_check}). Most unassociated picks from both models are isolated detections with no coincident pick elsewhere in the network, but PhaseNet+ produces 2.8 times as many isolated detections as EQNet2. Over the full day, EQNet2 locates 76 events, compared with 27 for PhaseNet+ and 3 in the USGS catalog.
The relative advantage of EQNet2 also persists in a test with more distant stations around the $M_w\,7.1$ Ridgecrest mainshock (\suppfigref{fig:case_ridgecrest}). Although both models degrade with increasing source--station distance, EQNet2 retains more arrivals that align across the network.

\subsection{Learned attention across stations follows seismic moveout}

\begin{figure}[htbp]
    \centering
    \includegraphics[width=\linewidth]{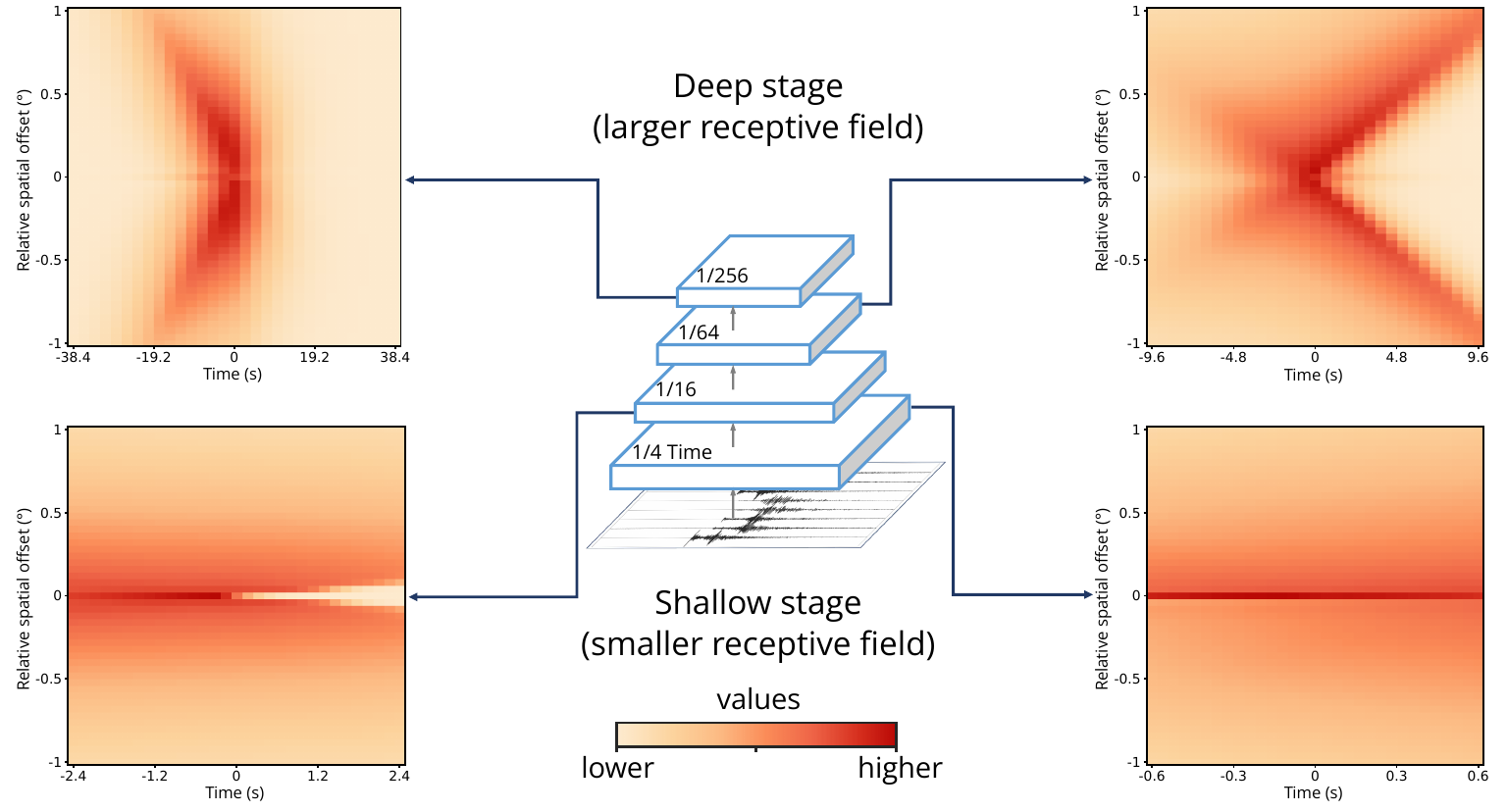}
    \caption{\textbf{Relative position biases across the EQNet2 backbone.} One attention head is shown for each of the four Swin Transformer stages. The horizontal axes denote relative time between waveform tokens, and the vertical axes denote the signed spatial offset of one station relative to the other, in degrees. Because the learned biases are nearly rotationally symmetric, the inter-station azimuth is fixed in the north--south direction and the relative elevation at zero. Darker colors indicate larger bias values. More visualizations are provided in \suppfigtworef{fig:rpb_heads_shallow_mid}{fig:rpb_heads_deep}.}
    \label{fig:rpb_first_block}
\end{figure}

The preceding tests establish the empirical benefits of network information. We next examine the station-to-station interactions learned by EQNet2.
The learned relative position biases (RPBs), which bias the attention between two waveform segments according to their separation in time and space, change systematically with model depth (Figure~\ref{fig:rpb_first_block}; \suppfigtworef{fig:rpb_heads_shallow_mid}{fig:rpb_heads_deep}). In shallow stages, high biases remain concentrated near small inter-station distances, emphasizing local information from the same or adjacent stations. In deeper stages, the high-bias patterns extend toward more distant stations and become increasingly slanted as the time span visible to each token grows. The slopes of some RPB heads correspond to apparent velocities of approximately 6--8~km/s, indicating that the model preferentially connects waveform segments whose time offsets match plausible travel times between stations.

\begin{figure}[htbp]
    \centering
    \includegraphics[width=\linewidth]{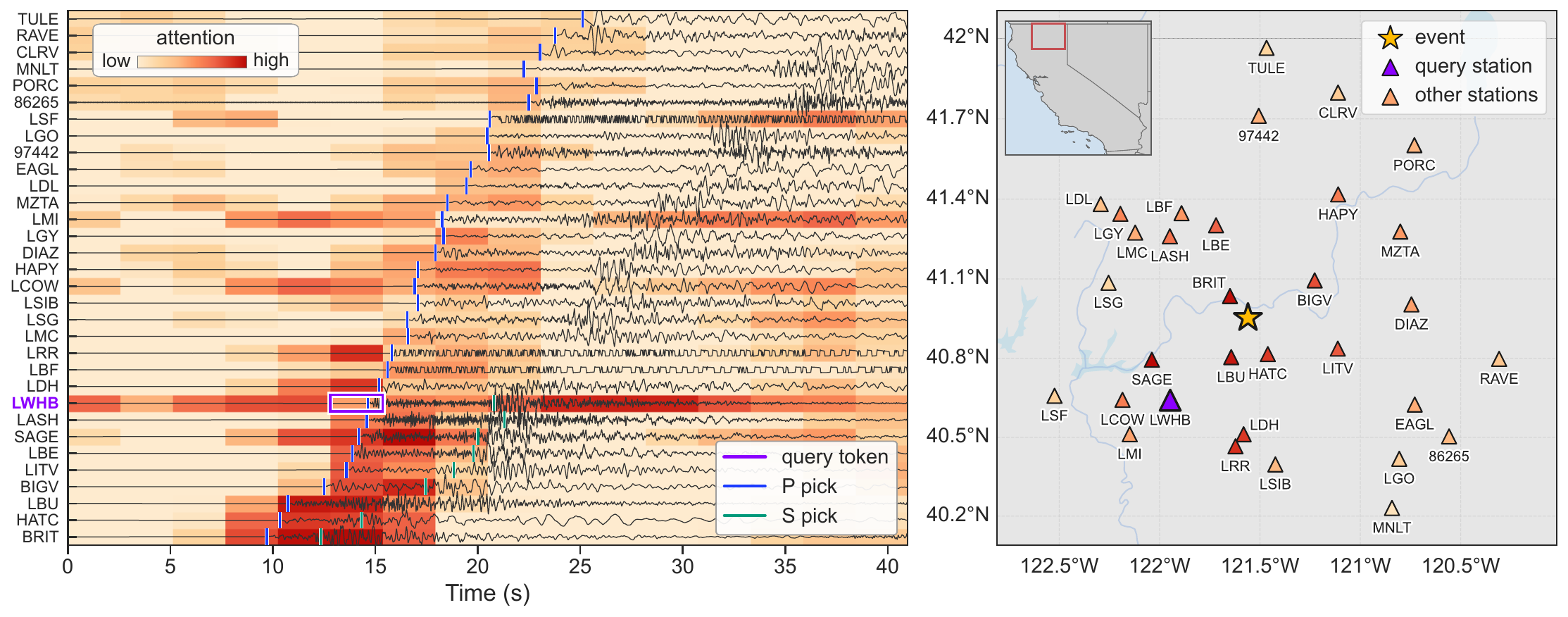}
    \caption{\textbf{Cross-station attention for a representative event.} The attention map for event nc73934641 from the last Swin Transformer block is shown together with the corresponding station distribution. The query token containing a P-phase arrival is enclosed by a purple rectangle, the query station is marked by a purple triangle, and darker colors indicate higher attention weights.}
    \label{fig:attn_map_last_block}
\end{figure}

Attention maps provide complementary event-specific evidence (Figure~\ref{fig:attn_map_last_block}; \suppfigref{fig:attention_maps_additional}). When the model queries a segment containing a P arrival, it assigns high attention to P arrivals at other stations, particularly at nearby stations or stations with similar waveforms. These patterns recur across query stations, indicating that the model links arrivals sharing a common moveout rather than treating each station independently.

\section{Discussion and Conclusions}

EQNet2 performs the tasks of earthquake monitoring in a single model that learns from multiple stations jointly and organizes its predictions around individual earthquakes.
In a fixed-window benchmark on Northern California data excluded from training, EQNet2 improves the picking of low-SNR arrivals. On continuous waveforms from the 2019 Ridgecrest sequence and the 2025 Berkeley earthquake, these gains translate into more recovered events, fewer false detections, and a substantially larger fraction of picks retained after association and location. The learned relative position biases form time--distance patterns consistent with seismic moveout, and the attention maps connect the arrivals of the same event across stations. These results show that combining network information with event-centric prediction improves both the completeness and the reliability of the catalog.

Multi-station modeling helps mainly by distinguishing weak arrivals that are coherent across the network from noise that is confined to a single station. A low-SNR arrival may be ambiguous at a single station but gains support from its consistent timing across the network. Conversely, cultural or instrumental noise can mimic a phase at one station but shows no consistent moveout across the network. The time--distance patterns in the learned relative position biases and the attention maps that connect the arrivals of the same event across stations indicate that the model has learned to exploit this coherence. Multi-station models have shown that network information can improve phase picking \citep{chen2022cubenet,feng2022edgephase,sun2023phaseno}. EQNet2 extends this by conditioning the attention between stations on their relative geometry throughout the shared backbone, so that the geometric relations between stations inform the picking, source, and association predictions alike.

Event-centric prediction contributes largely by optimizing the tasks jointly rather than in sequence, on a shared representation of each earthquake. Phase arrivals, first-motion polarities, event-arrival proxies, source-related estimates, and association masks are different measurements of the same earthquake. Predicting them from a shared representation lets the evidence for one measurement support the other measurements and encourages the outputs to be mutually compatible. In a sequential workflow, by contrast, independently optimized modules communicate primarily through discrete intermediate predictions, which discard the waveform evidence and its uncertainty and let early errors propagate. For example, the prompt-based association module takes a candidate event-arrival proxy as the prompt and predicts a network-wide association probability mask rather than treating association as a pairwise matching task \citep{ross2019phaselink,mcbrearty2019pairwise}. Multi-task models have shown that related outputs can benefit from joint learning \citep{zhu_eqnet_2022,zhang2023simultaneous,si2024plan,zhu2025towards}. EQNet2 extends this by organizing all of its outputs around candidate events, so that the station-level measurements and the network-wide association and prediction refer to the same earthquake.

The computational efficiency of EQNet2 balances multi-station modeling, which increases the cost of each prediction, against event-centric prediction, which reduces the complexity of the whole monitoring workflow. Exchanging information across stations adds computation compared with processing each station independently, while the resulting features are shared across tasks and lead to more reliable detections. Event-centric prediction in turn reduces the complexity of the workflow by replacing several separate models and intermediate stages with one shared backbone and parallel task heads. EQNet2 is thus more expensive than lightweight single-station models but faster and more memory-efficient than other multi-station models such as PhaseNO under the same benchmark setting (\supptabref{tab:compute_cost}).
In the Ridgecrest test, processing one day of continuous waveforms from 45 sensors takes about 10 minutes on a single NVIDIA A100 GPU, and post-processing all six days of output takes a few additional minutes.
Model capacity and station-window size provide practical controls on the performance--efficiency trade-off. Smaller EQNet2 variants reduce runtime with a limited decrease in phase-picking performance on the NCEDC test dataset (\supptabref{tab:compute_cost}), whereas larger variants produce fewer raw picks and retain a larger fraction of them after association and location in continuous data (\supptabref{tab:association_ratio_scaling}). Enlarging the station window includes more stations in the attention for each event, which reduces split events and raises the fraction of picks retained (\suppfigref{fig:ridgecrest_raw_location}; \supptabref{tab:association_ratio_scaling}), at the cost of higher memory use.

Several limitations of the present model point to future development. First, event-centric training depends on datasets in which the phase, polarity, source, and association labels are consistent for each event, and on augmentation that preserves this consistency; larger and more diverse datasets of this kind would improve generalization. Second, multi-task optimization remains challenging because picking, polarity, source-related estimation, and association losses can have different scales and convergence rates \citep{liu2019mtl}. Third, applying the model to larger networks will require sparser station selection or hierarchical station windows to keep the computational cost manageable. In addition, some RPB heads are asymmetric in time (Figure~\ref{fig:rpb_first_block}), which we cannot yet explain physically. Understanding these learned patterns may help assess the reliability and generalization of the model.

This study shows that combining network coherence with event-centric prediction can recover low-SNR arrivals and suppress false detections, yielding more complete and more reliable catalogs. The attention learned across the network follows the physics of seismic wave propagation, suggesting that the model improves its predictions through a physically interpretable representation of the travel times between stations. The gains in completeness and reliability support moving from station-by-station and step-by-step workflows toward models that combine geometry-aware attention across stations with event-centric prediction. Future work will focus on generalization across regions and network geometries and on scaling to larger and denser seismic networks, where the coherence across closely spaced stations could yield catalogs that resolve fault structure and earthquake processes in finer detail.

\section*{Data Availability}

The CEED dataset used for training is available at Hugging Face \citep{zhu2025california}. The continuous seismic data for the 2019 Ridgecrest earthquake sequence, the waveform-relocated SCSN catalog, and the refined focal mechanism solutions by \citet{cheng2023refineda} were downloaded from \citet{scedc2013}. The continuous seismic data for the 2025 Berkeley earthquake were downloaded from \citet{ncedc2014}.

\section*{Code Availability}

Our source code and model are publicly available at \url{https://github.com/AI4EPS/EQNet}.

\section*{Acknowledgements}

This work was supported by the U.S. Department of Energy, Office of Science, Basic Energy Sciences, under Award DE-SC0026120. This research used the Lawrencium computational cluster provided by the IT Division at Lawrence Berkeley National Laboratory (supported by the Director, Office of Science, Office of Basic Energy Sciences, of the U.S. Department of Energy under Contract No. DE-AC02-05CH11231). This research used resources of the National Energy Research Scientific Computing Center (NERSC), a Department of Energy User Facility (project m4781-2024). We thank Bo Rong, Xuancheng Huang, Junhao Song, and Savvas Marcou for their valuable suggestions on paper writing and figure style.

\bibliography{references}
\clearpage
\section*{Supplementary Information}
\setcounter{figure}{0}
\setcounter{table}{0}
\setcounter{equation}{0}
\renewcommand{\thefigure}{\arabic{figure}}
\renewcommand{\thetable}{\arabic{table}}
\renewcommand{\theequation}{\arabic{equation}}
\renewcommand{\figurename}{Supplementary Fig.}
\renewcommand{\tablename}{Supplementary Table}

\newcounter{suppnote}
\newcommand{\suppnote}[2]{%
    \refstepcounter{suppnote}%
    \section*{Supplementary Note \arabic{suppnote}: #1}%
    \label{#2}%
}

\suppnote{Details of model structure}{sec:supp_model}

Input waveforms are normalized and divided into short time segments (tokens). These tokens are then processed by a series of Swin Transformer blocks with progressive temporal downsampling, producing a hierarchy of multi-scale features. The key mechanism within these blocks is shifted window attention. Window attention is computed within local windows spanning both time and stations to reduce cost relative to global attention, and shifting the windows between successive blocks lets information flow across window boundaries. This design suits large seismic networks, where attention over all stations would be costly and the relevant information is expected to come mainly from nearby stations with similar waveforms.

The multi-scale backbone features are fused by an FPN module \citep{lin2017feature,xiao_upernet}. Since the backbone already models cross-station interactions through attention, the FPN and most task-specific heads primarily operate along the time dimension using one-dimensional convolutional modules.

The task heads are shown in Supplementary Fig.~\ref{fig:heads_arch}. The Phase, Polarity, and Event Heads consist mainly of 1D convolutional layers. To improve temporal precision, the Polarity Head additionally uses shallow high-resolution features from the backbone. The source-related estimates, such as travel time and epicentral distance, are read from the regression output sequences at the predicted event-arrival proxy time. Station-wise origin times are then obtained by subtracting the predicted travel time from the proxy time, and station-wise epicenters by projecting the predicted epicentral distance along the predicted backazimuth from each station.

The Association Head uses a prompt-based attention mechanism to predict association masks from backbone features. During inference, the Association Head is applied to all detected arrivals. To convert these masks into candidate events, we construct a graph whose nodes are detected event-arrival proxies, and edges represent association probabilities. Each connected component of the graph is treated as one candidate event, which merges queries belonging to the same earthquake.

\begin{figure}[htbp]
    \centering
    \includegraphics[width=0.8\linewidth]{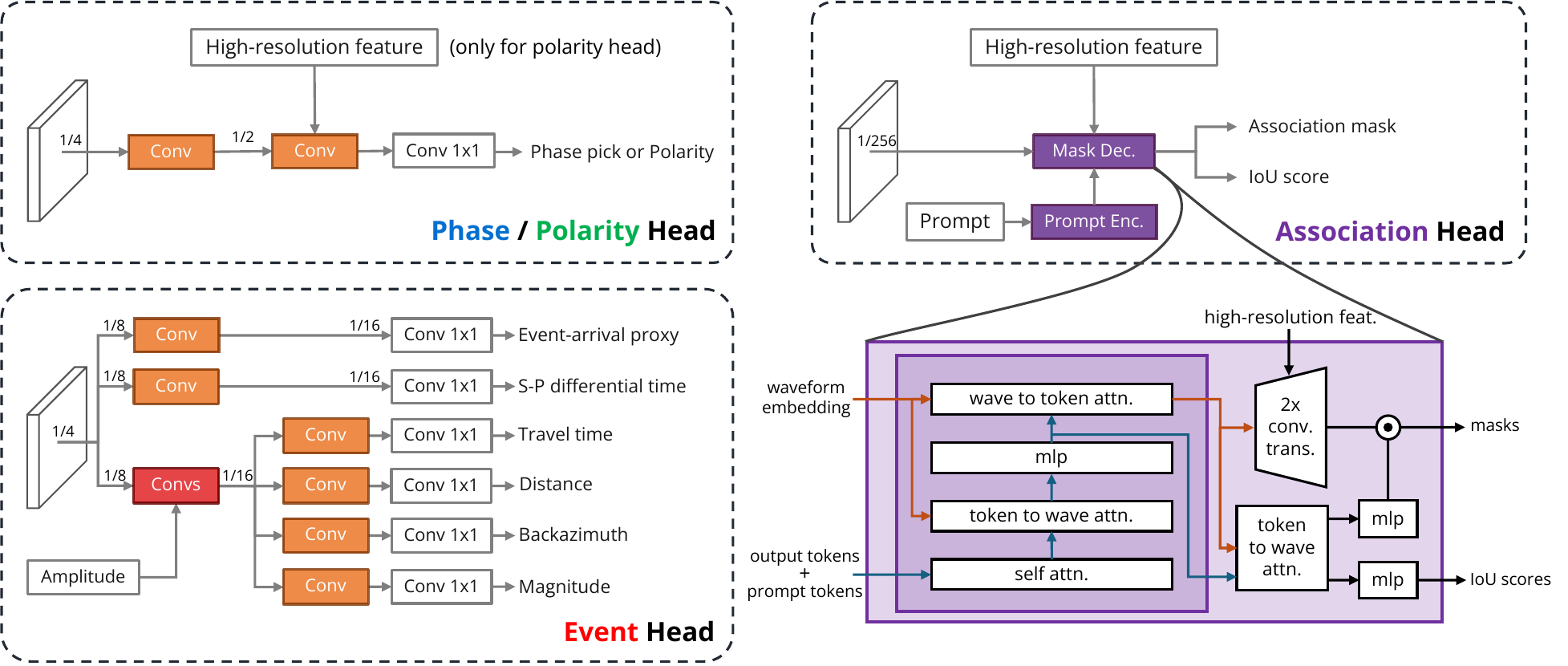}
    \caption{Structure of task heads. The Association Head is modified from the Segment Anything Model \citep{kirillov2023segment,ravi2024sam}; the other heads mainly consist of 1D convolution modules.}
    \label{fig:heads_arch}
\end{figure}

\suppnote{Ordinal regression for polarity picking}{sec:ordinal_regression}

The first-motion polarity can be classified into three classes: Down, Unknown, and Up, which have an inherent order. Because the three classes are ordered, we formulate polarity picking as ordinal regression and convert the model output into class likelihoods \citep{mccullagh1980regression}.

For a binary classification problem, the likelihoods of classes 0 and 1 can be $p_0 = 1 - \sigma(x)$ and $p_1 = \sigma(x)$, where $\sigma(x) = 1/(1+e^{-x})$ is the sigmoid function and $x$ is the predicted logit. For a general $k$-class classification, the problem can be solved by extending the sigmoid function to the softmax function and the scalar logit to a $k$-dimensional logit. However, for ordinal problems, we can interpret the binary case as $p_0 = p(x\leq0) = \sigma(0-x)$ and $p_1 = p(x>0) = 1 - p(x\leq0)$. For a more general case, with cut points \{$c_0=-\infty$, $c_1$, $c_2=\infty$\}, binary classification can be described as

\begin{equation}
    \begin{aligned}
        p_0 &= p(c_0 \leq x\leq c_1) = \sigma(c_1-x) - \sigma(c_0-x) = \sigma(c_1-x) \\
        p_1 &= p(c_1<x\leq c_2) = \sigma(c_2-x) - \sigma(c_1-x) = 1 - \sigma(c_1-x)
    \end{aligned}
\end{equation}

Extending to the multinomial case, the likelihood of ordinal class $k$ can be defined as

\begin{equation}
    p_k = p(c_{k-1}<x\leq c_k) = \sigma(c_k-x) - \sigma(c_{k-1}-x)
\end{equation}

The model is optimized by minimizing the negative log-likelihood of these class probabilities. For polarity picking, $x$ is the output logit, the classes are ordered as Down, Unknown, and Up, and the cut points are $c_1=-\theta$ and $c_2=\theta$, with $c_0=-\infty$ and $c_3=\infty$, where $\theta$ is a hyperparameter.

\suppnote{Loss functions and training details}{sec:supp_losses}

The total loss is a weighted sum of task losses:
\begin{equation}
    \mathcal{L} = \lambda_{\text{phase}}\mathcal{L}_{\text{phase}} +
                 \lambda_{\text{event}}\mathcal{L}_{\text{event}} +
                 \lambda_{\text{pol}}\mathcal{L}_{\text{pol}} +
                 \lambda_{\text{reg}}\mathcal{L}_{\text{reg}} +
                 \lambda_{\text{asso}}\mathcal{L}_{\text{asso}}
\end{equation}

It includes cross-entropy for phase picking and event-arrival-proxy detection, a combination of focal \citep{lin2017focal} and dice loss \citep{sudre2017generalised} for association, negative log-likelihood for polarity determination, and smooth L1 loss for regression tasks. To handle data imbalance, additional weighting is applied to regression targets with uneven distributions, such as depth and magnitude. Rarer target values, including larger magnitudes, are assigned higher weights in the corresponding loss terms. We also apply masks to polarity and regression losses so that only relevant event-related time windows contribute to optimization. Because earthquake catalogs are incomplete and some true arrivals may be unlabeled, we avoid treating all unlabeled samples as negative examples. For phase picking and event-arrival detection, masked loss regions are used to reduce the penalty on potentially unlabeled arrivals. We also randomly sample points from noise windows to balance the ratio of positive and negative samples during training.

We employ the AdamW optimizer \citep{loshchilov2017decoupled} with a base learning rate of 0.001, a weight decay of 0.05, and a batch size of 40, with a cosine decay learning rate scheduler and a one-epoch warm-up. Training the model for 33 epochs takes approximately 24 hours using 4 NVIDIA A100 GPUs, depending on data-loading efficiency.

We use two data augmentation strategies. The first is temporal event stacking, which is widely used in single-station model training \citep{zhu2020seismic}. We adapt the network-level stacking strategy used in PhaseNO \citep{sun2023phaseno}.
For a given primary event, we randomly sample a secondary event sharing at least three stations, scale its waveforms by a random factor to simulate magnitude differences, adjust the magnitude label accordingly, apply a random time shift, and stack the scaled waveforms onto the primary event at shared stations.
This simulates temporally dense event distributions common in earthquake sequences.
The second strategy is spatial event concatenation, designed to simulate scenarios where signals are recorded by a subset of the network. For a given primary event, we randomly select a secondary event of similar magnitude and sample a random subset of stations from both.
The combined data are concatenated along the station dimension, creating inputs that encourage the Association Head to distinguish nearby but distinct events.

\suppnote{More metrics on the NCEDC 2023 dataset}{sec:metrics_vs_params}

We evaluate picking and detection performance using precision, recall, and F1 score. Precision measures the fraction of predicted positives that are correct, whereas recall measures the fraction of reference positives that are recovered:
\begin{equation}
    \mathrm{Precision} = \frac{\mathrm{TP}}{\mathrm{TP}+\mathrm{FP}},
    \qquad
    \mathrm{Recall} = \frac{\mathrm{TP}}{\mathrm{TP}+\mathrm{FN}}.
\end{equation}
The F1 score is the harmonic mean of precision and recall:
\begin{equation}
    \mathrm{F1} = \frac{2\times\mathrm{Precision} \times \mathrm{Recall}}
    {\mathrm{Precision}+\mathrm{Recall}}.
\end{equation}

For phase picking, a predicted phase arrival is counted as a true positive (TP) if it is matched to a reference arrival of the same phase type with an absolute time residual smaller than 0.5~s. For event detection, a predicted event is counted as a TP if the absolute origin-time residual is smaller than 3~s. Unmatched predictions are counted as false positives (FP), and unmatched reference arrivals or events are counted as false negatives (FN).

\begin{figure}[htbp]
    \centering
    \includegraphics[width=\linewidth]{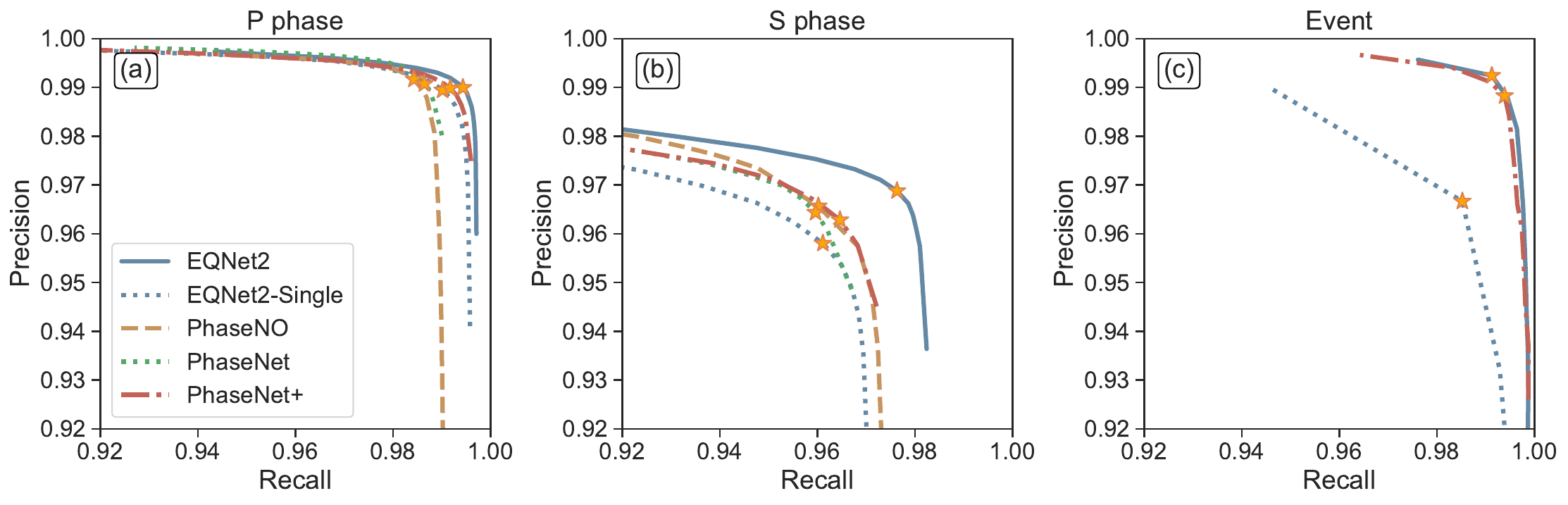}
    \caption{Precision--recall curves of phase picking and event detection on the NCEDC 2023 dataset. Yellow stars mark the highest F1 score for each model.}
    \label{fig:pr_curve}
\end{figure}

\begin{figure}[htbp]
    \centering
    \includegraphics[width=\linewidth]{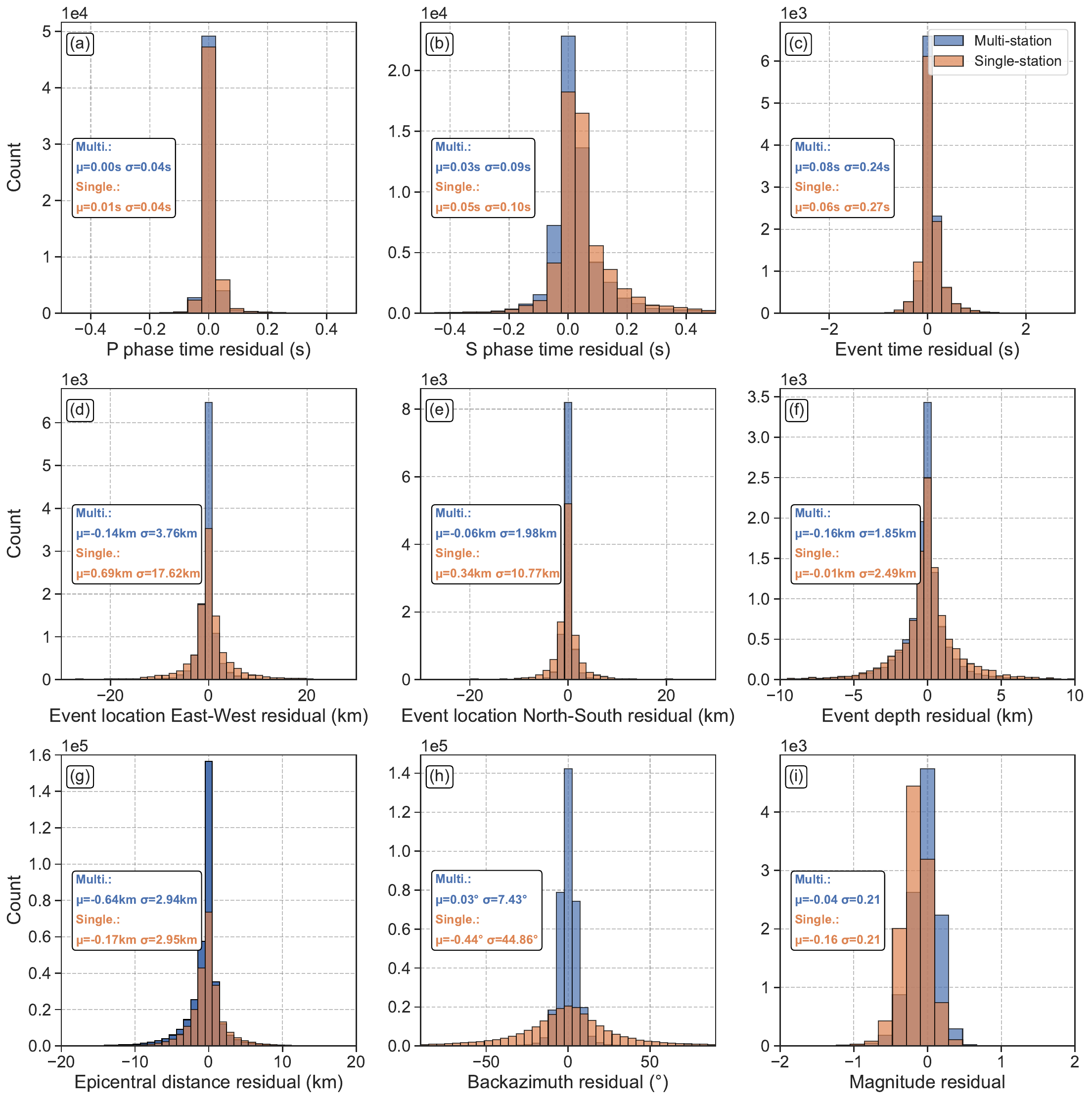}
    \caption{Phase picking and event detection residual distributions of EQNet2 and its single-station variant on the NCEDC 2023 dataset.}
    \label{fig:metrics_residuals}
\end{figure}

\begin{figure}[htbp]
    \centering
    \includegraphics[width=\linewidth]{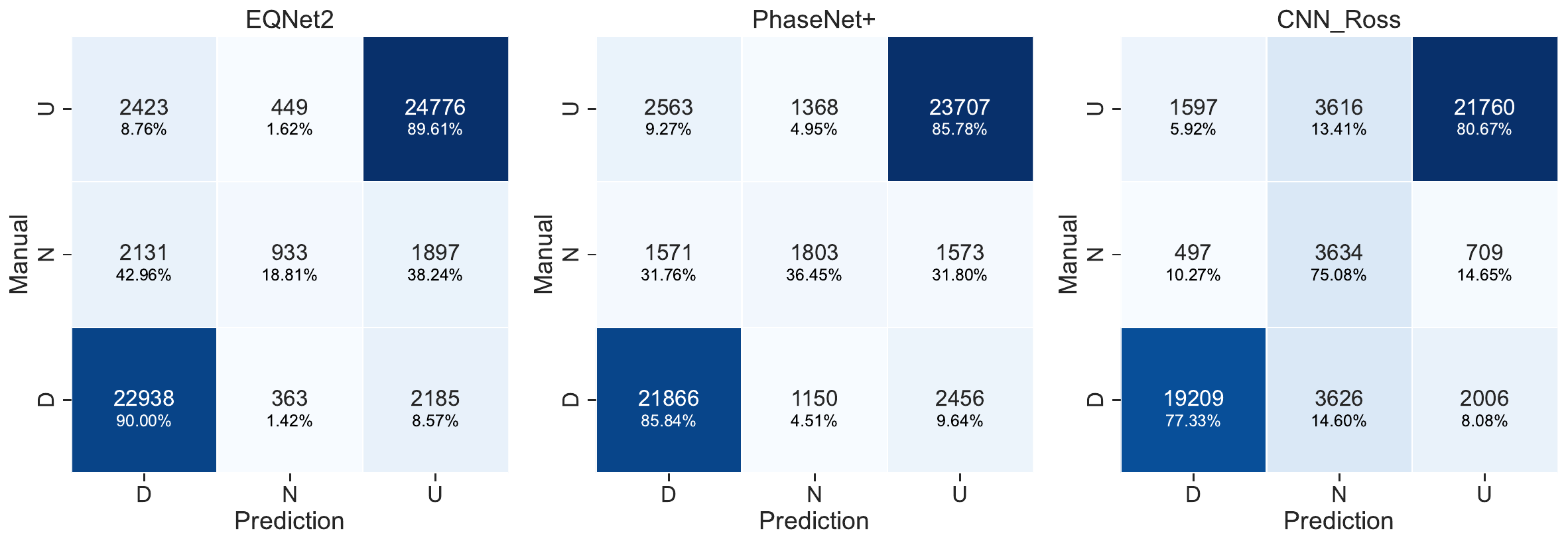}
    \caption{Confusion matrix for P-phase first-motion polarity determination on the NCEDC 2023 dataset. The three classes, D, N, and U, denote Down, Unknown, and Up polarities, respectively.}
    \label{fig:polarity_confusion_matrix}
\end{figure}

\begin{figure}[htbp]
    \centering
    \includegraphics[width=0.8\linewidth]{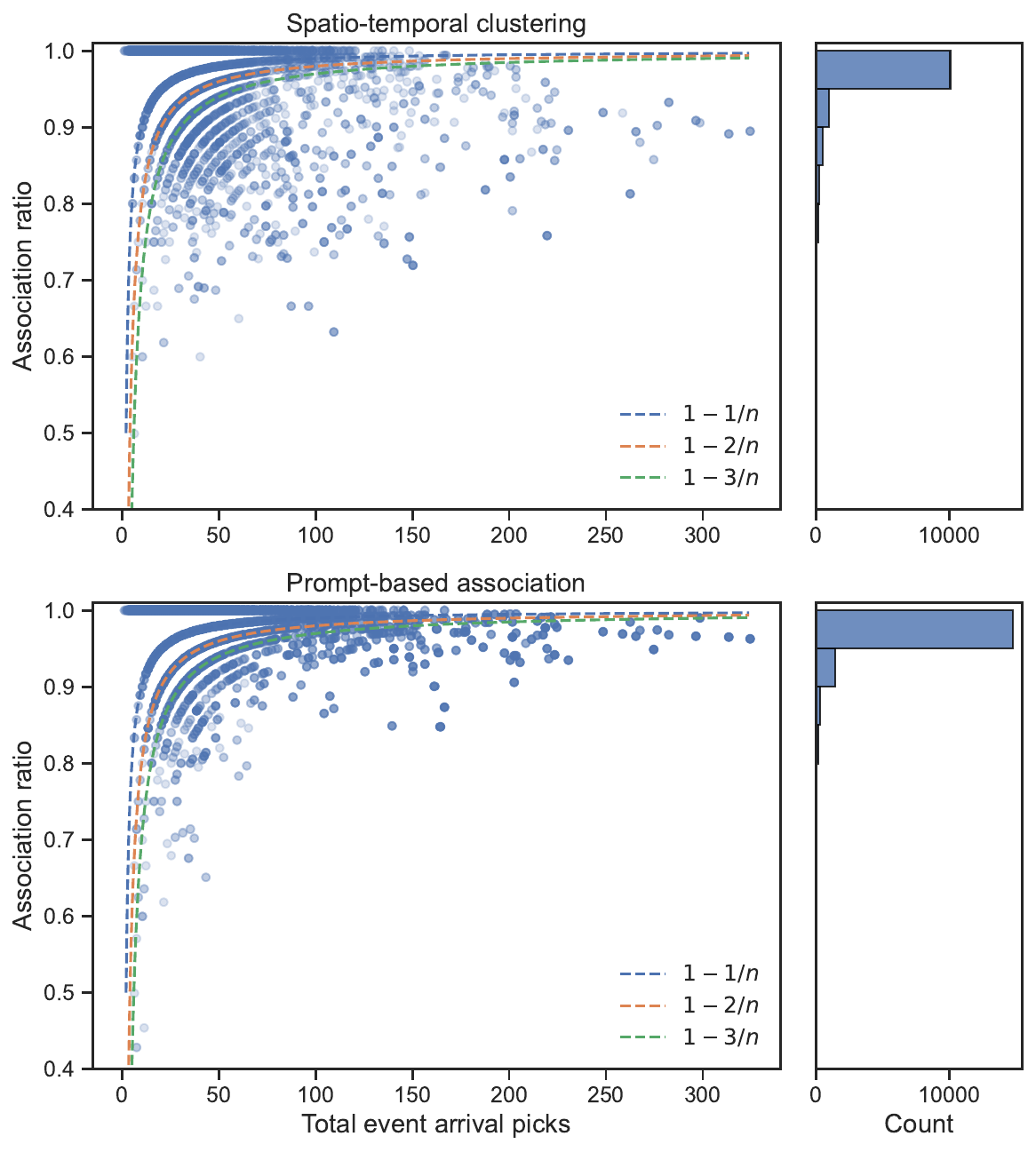}
    \caption{Window-wise association ratio on the NCEDC 2023 dataset by spatiotemporal clustering and prompt-based method. The association ratios are expected to concentrate on curves whose equations follow $1 - \frac{k}{n}$, where $n$ is the total number of signals and $k$ is the number of unassociated ones.}
    \label{fig:association_ratio}
\end{figure}

\FloatBarrier

\begin{table}[hbp]
  \centering
  \caption{Phase-picking metrics and time-residual statistics at a score threshold of 0.3.}
    \begin{tabular}{llSSSSS}
    \toprule
      {}  & {} & {Precision} & {Recall} & {F1}    & {Mean (s)} & {Std (s)} \\
    \midrule
    \multicolumn{1}{l}{EQNet2} & P-phase & 0.9824  & 0.9966  & 0.9895  & 0.004  & 0.038  \\
          & S-phase & 0.9610  & 0.9803  & 0.9705  & 0.027  & 0.088  \\
    \multicolumn{1}{l}{EQNet2-Single} & P-phase & 0.9793  & 0.9946  & 0.9869  & 0.007  & 0.039  \\
    \multicolumn{1}{l}{} & S-phase & 0.9487  & 0.9671  & 0.9578  & 0.049  & 0.101  \\
    \multicolumn{1}{l}{PhaseNO} & P-phase & 0.9419  & 0.9899  & 0.9653  & 0.003  & 0.050  \\
          & S-phase & 0.9366  & 0.9724  & 0.9542  & 0.010  & 0.100  \\
    \multicolumn{1}{l}{PhaseNet} & P-phase & 0.9907  & 0.9864  & 0.9886  & -0.001  & 0.039  \\
          & S-phase & 0.9672  & 0.9567  & 0.9619  & 0.016  & 0.103  \\
    \multicolumn{1}{l}{PhaseNet+} & P-phase & 0.9884  & 0.9930  & 0.9907  & -0.001  & 0.038  \\
          & S-phase & 0.9670  & 0.9592  & 0.9631  & 0.011  & 0.100  \\
    \bottomrule
    \end{tabular}%
  \label{tab:phase_picking_metric}%
\end{table}%

\begin{table}[htbp]
\centering
\caption{Event detection residual statistics at threshold 0.3.}
\label{tab:event_detection_metric}

\begin{threeparttable}
\small
\setlength{\tabcolsep}{5pt}

\begin{tabular}{
lSSSSSSSSS
}
\toprule

& \multicolumn{3}{c}{Detection Metrics}
& \multicolumn{2}{c}{Origin Time (s)}
& \multicolumn{2}{c}{Distance (km)}
& \multicolumn{2}{c}{Backazimuth (deg)} \\

\cmidrule(lr){2-4}
\cmidrule(lr){5-6}
\cmidrule(lr){7-8}
\cmidrule(lr){9-10}

Model & {Precision} & {Recall} & {F1} & {Mean} & {Std} & {Mean} & {Std} & {Mean} & {Std} \\

\midrule

EQNet2
& 0.9106
& 0.9986
& 0.9525
& 0.075
& 0.243
& -0.638
& 2.941
& 0.027
& 7.428 \\

EQNet2-Single
& 0.7255
& 0.9971
& 0.8399
& 0.062
& 0.278
& -0.174
& 2.954
& -0.443
& 44.863 \\

PhaseNet+
& \bfseries 0.9260
& \bfseries 0.9987
& \bfseries 0.9610
& \bfseries 0.004
& 0.264
& {--}
& {--}
& {--}
& {--} \\

\bottomrule
\end{tabular}

\end{threeparttable}
\end{table}

\FloatBarrier

\begin{figure}[htbp]
    \centering
    \includegraphics[width=0.7\linewidth]{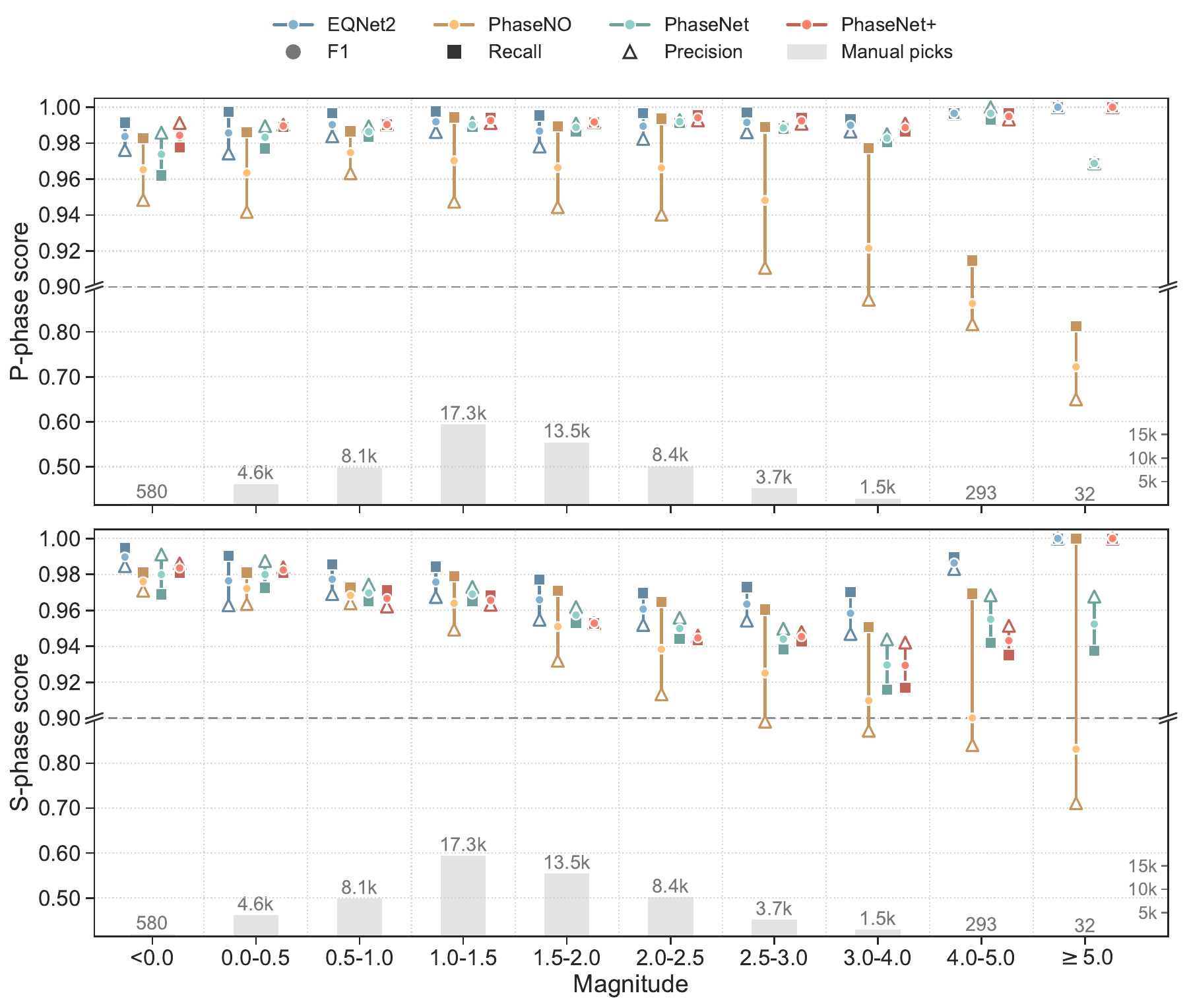}
    \caption{Phase picking F1, precision and recall scores for P- and S-phase arrival picking on the NCEDC 2023 dataset across magnitude groups. Symbols are the same as Figure~\ref{fig:metrics_vs_snr_f1} in the main text.}
    \label{fig:metrics_vs_magnitude_f1}
\end{figure}

\begin{figure}[htbp]
    \centering
    \includegraphics[width=0.7\linewidth]{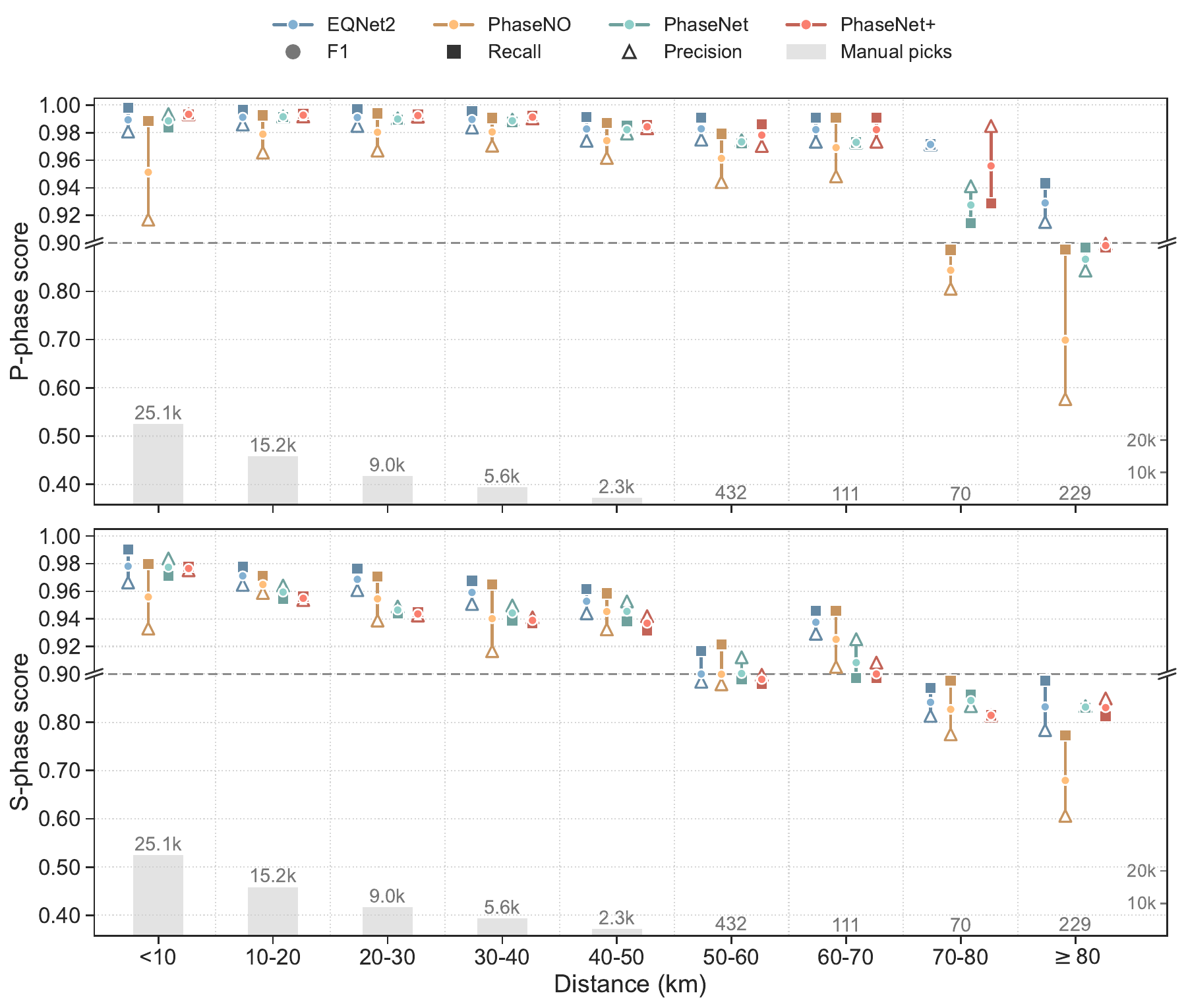}
    \caption{Phase picking F1, precision and recall scores for P- and S-phase arrival picking on the NCEDC 2023 dataset across epicentral-distance groups. Symbols are the same as Figure~\ref{fig:metrics_vs_snr_f1} in the main text.}
    \label{fig:metrics_vs_distance_f1}
\end{figure}

\begin{figure}[htbp]
    \centering
    \includegraphics[width=0.7\linewidth]{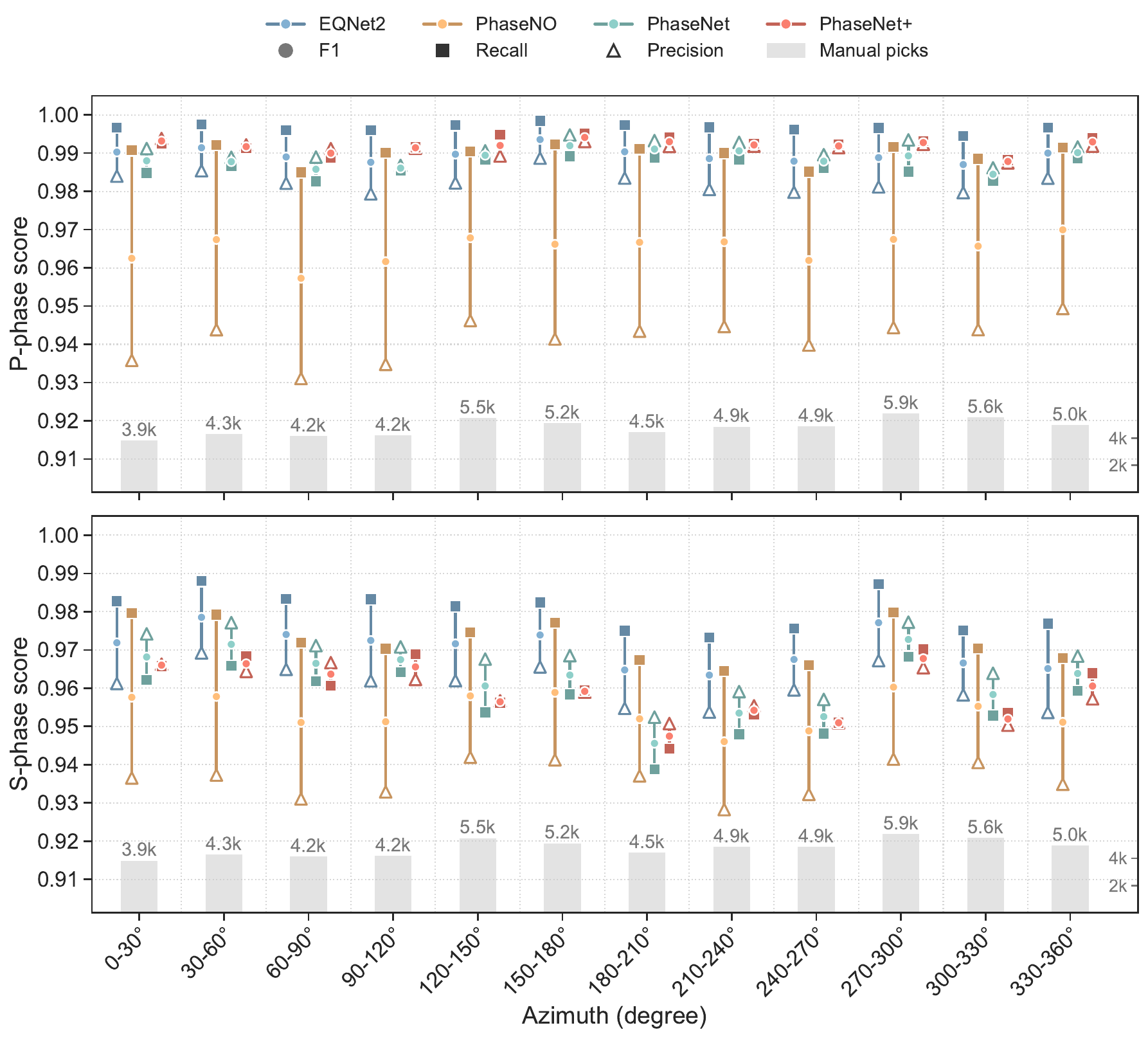}
    \caption{Phase picking F1, precision and recall scores for P- and S-phase arrival picking on the NCEDC 2023 dataset across azimuth groups. Symbols are the same as Figure~\ref{fig:metrics_vs_snr_f1} in the main text.}
    \label{fig:metrics_vs_azimuth_f1}
\end{figure}

\begin{figure}[htbp]
    \centering
    \includegraphics[width=1.0\linewidth]{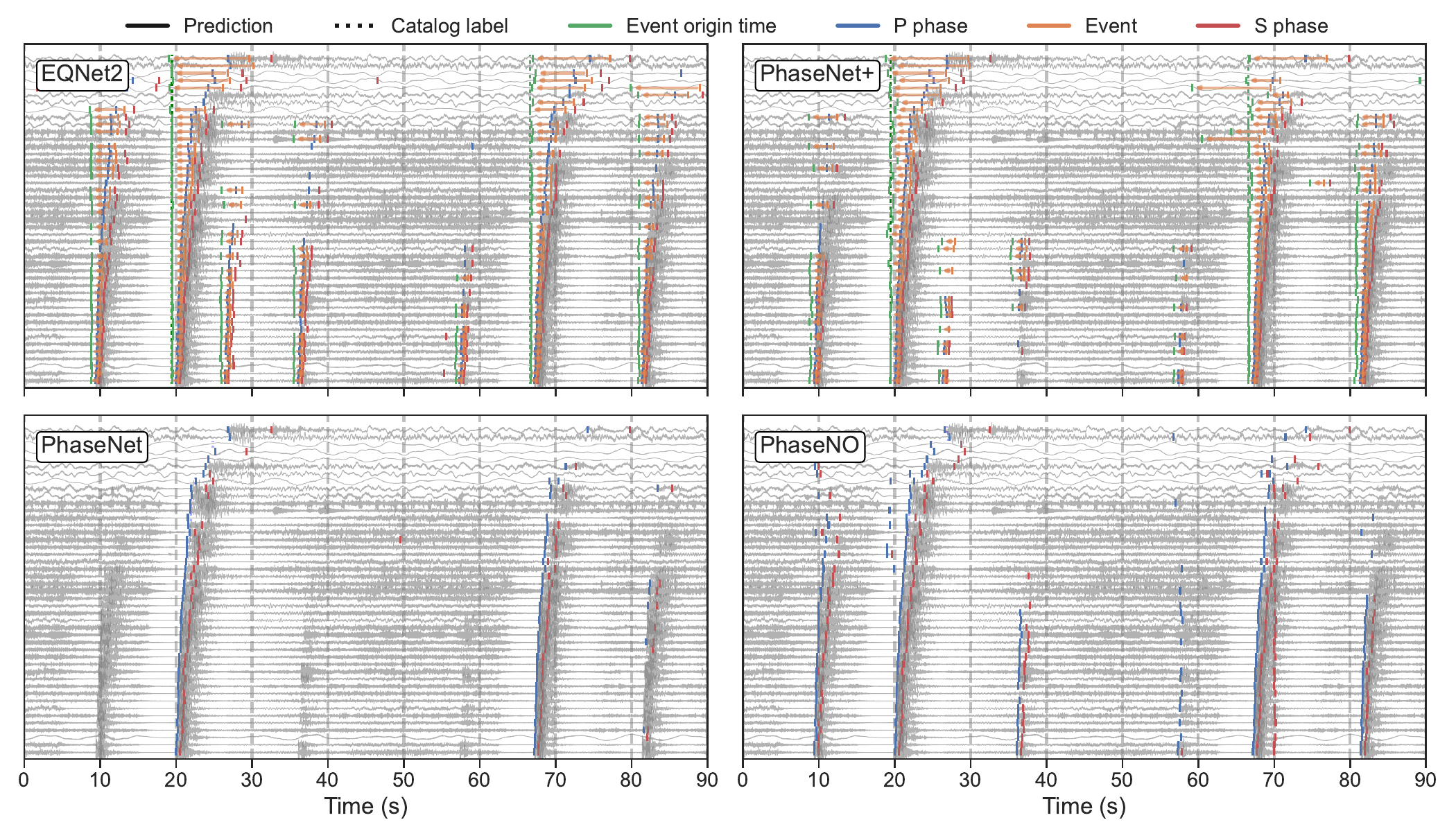}
    \caption{A dense-event window from the NCEDC 2023 test dataset. Predicted P- and S-phase arrivals from EQNet2, PhaseNet+, PhaseNet, and PhaseNO are shown for a 90-s window containing two cataloged events, nc73848531 and nc73848536. Solid lines denote predictions and dotted lines denote catalog annotations; blue and red indicate P- and S-phase arrivals, green indicates origin-time estimates, and orange indicates event-arrival proxies. Event-level predictions are available only for EQNet2 and PhaseNet+.}
    \label{fig:compare_test_dataset}
\end{figure}

\FloatBarrier

\suppnote{Additional results for the 2019 Ridgecrest earthquake sequence}{sec:ridgecrest_supp}

This note provides additional Ridgecrest results: preliminary locations from the model outputs before absolute location, focal-mechanism comparisons, and a pick-level comparison around the $M_w\,7.1$ mainshock.

\begin{figure}[htbp]
    \centering
    \includegraphics[width=\linewidth]{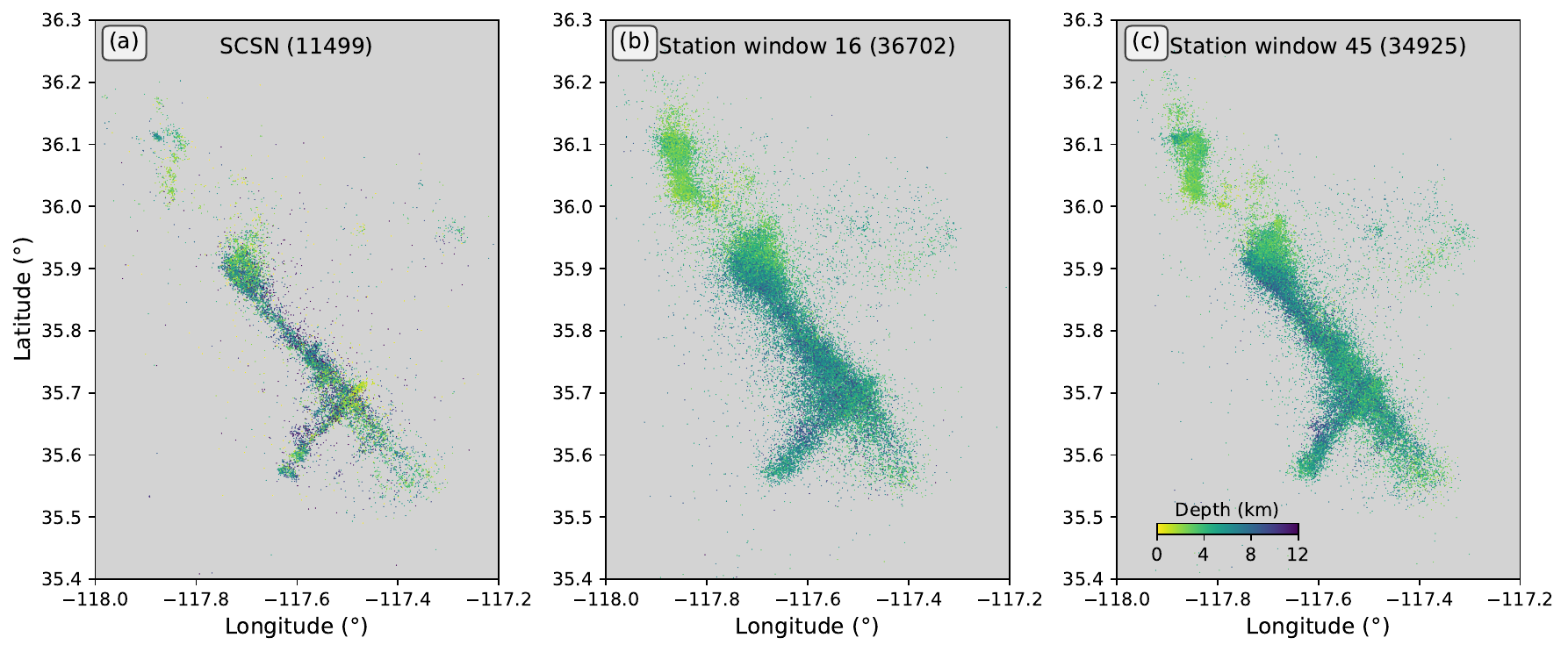}
    \caption{Epicenter locations predicted by EQNet2 with different station window sizes before the absolute location procedure: (a)~SCSN (reference), (b)~16, (c)~45 (all sensors).}
    \label{fig:ridgecrest_raw_location}
\end{figure}

\begin{figure}[htbp]
    \centering
    \includegraphics[width=\linewidth]{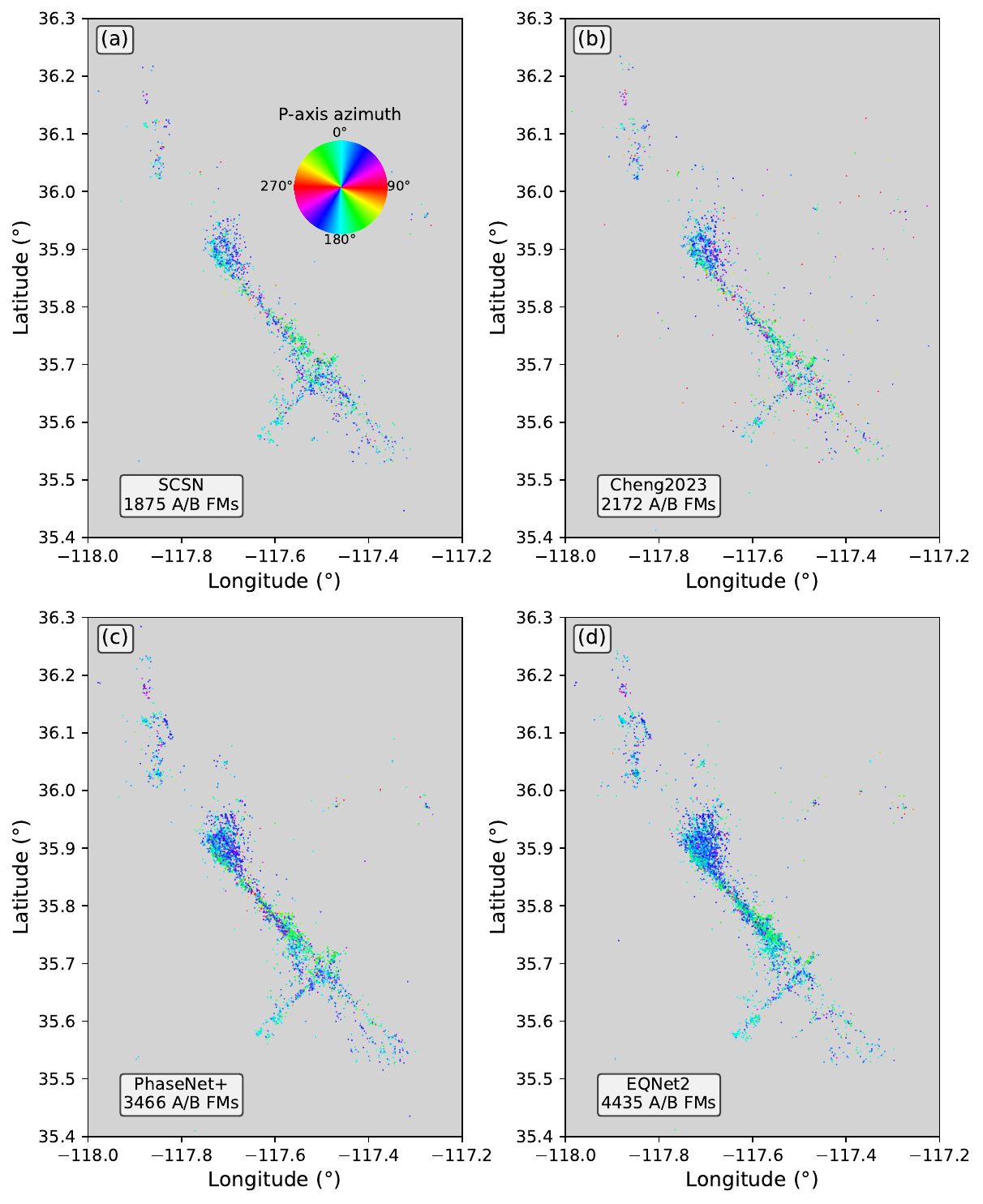}
    \caption{Spatial distribution of P-axis azimuths from focal mechanism solutions: (a) SCSN catalog, (b)
    catalog of \citet{cheng2023refineda}, (c) PhaseNet+ catalog, and (d) EQNet2 catalog.}
    \label{fig:ridgecrest_fm}
\end{figure}

\begin{figure}[htbp]
    \centering
    \includegraphics[width=\linewidth]{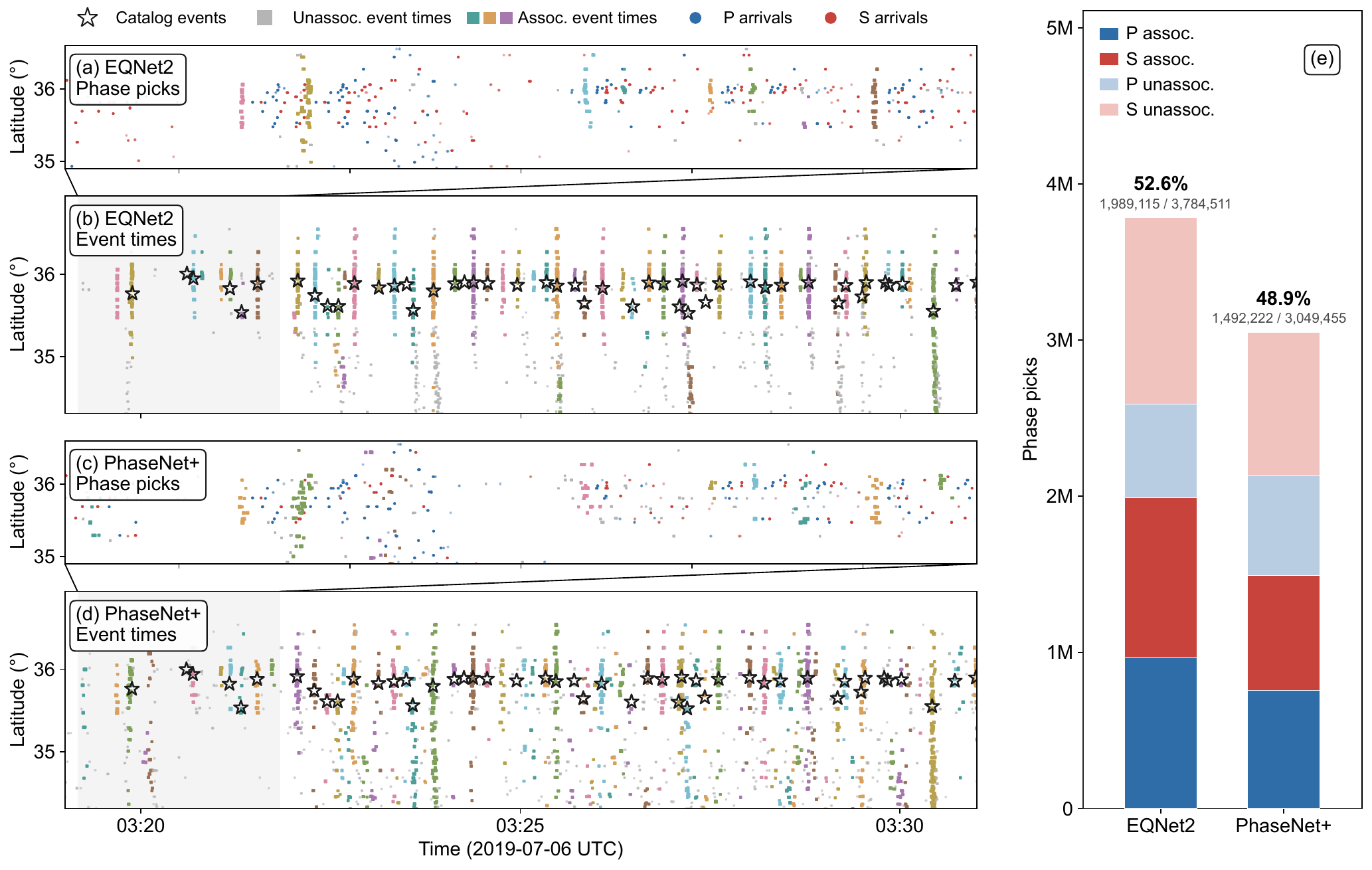}
    \caption{Comparison of EQNet2 and PhaseNet+ for the 2019 Ridgecrest earthquake sequence. (a--d), Predictions in a time window selected from the 2019 Ridgecrest sequence around the $M_w\,7.1$ mainshock (event ID: ci38457511). Each square in panels b and d denotes a predicted event origin time at the corresponding station. Gray squares are unassociated predictions, colored squares indicate predictions associated with the same event, and the black star marks the SCSN catalog event. Panels a and c enlarge the shaded interval and additionally show predicted P- and S-phase arrivals. (e) Stacked counts of associated and unassociated P- and S-phase picks. Percentages indicate the fraction of raw picks retained after association and location.}
    \label{fig:case_ridgecrest}
\end{figure}

\FloatBarrier

\suppnote{Examination of Berkeley earthquake detections}{sec:berkeley_check}

\begin{figure*}[htbp]
    \centering
    \includegraphics[width=\textwidth]{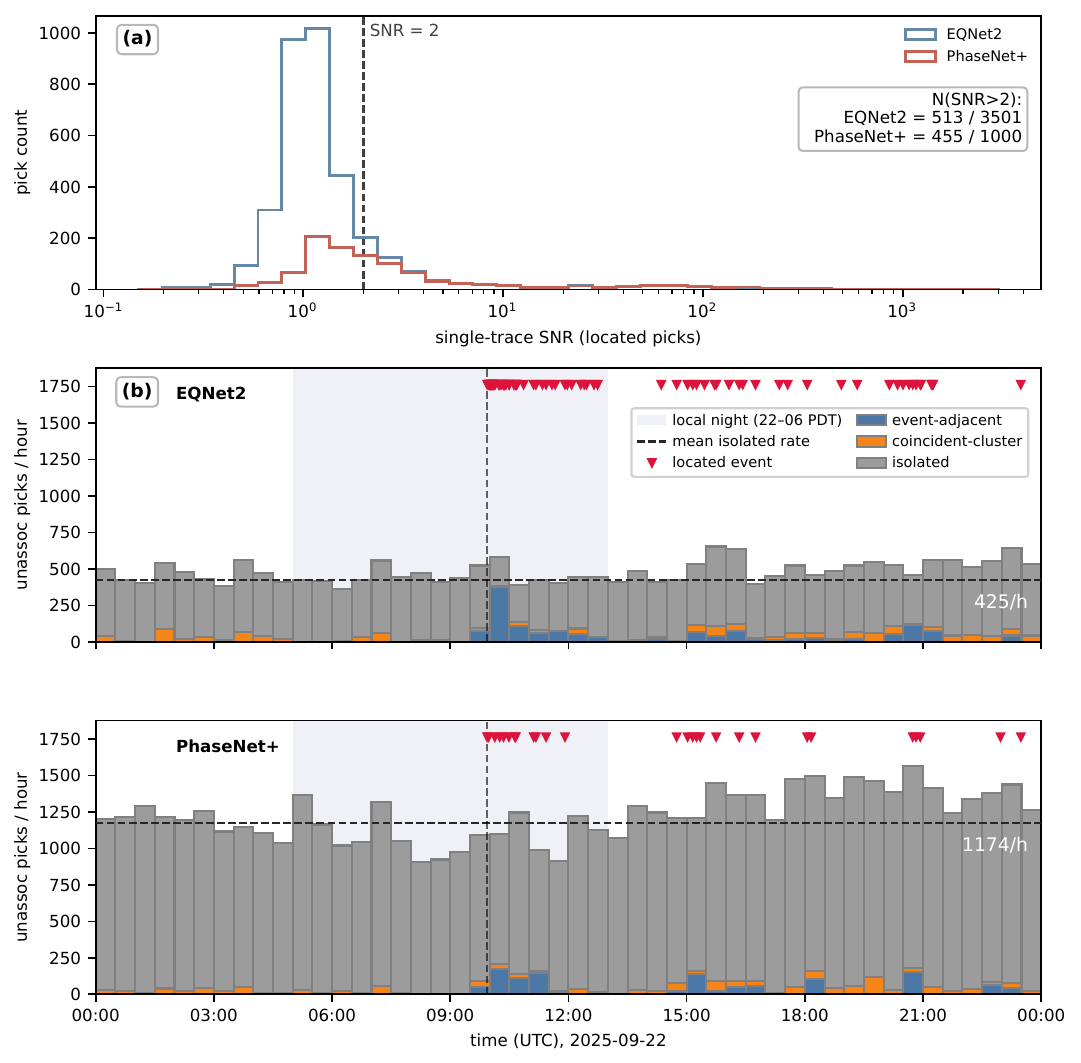}
    \caption{Pick-level evidence for the Berkeley one-day comparison. (a)~Counts of located picks as a function of single-trace signal-to-noise ratio for EQNet2 and PhaseNet+. (b)~Unassociated picks through the day, separated into picks adjacent to a located event, picks coincident across stations, and isolated picks.}
    \label{fig:sfig_picks}
\end{figure*}

Throughout, we refer to picks output by the models as \emph{raw}, picks kept by the associator as \emph{associated}, and picks used in the final location as \emph{located}, and we verify the events each model locates. Among the final located picks, the two models pick a similar number of strong arrivals above an SNR of 2 (513 for EQNet2 and 455 for PhaseNet+), while EQNet2 locates roughly 2,500 more low-SNR arrivals; each coincides with a median of 40 same-model picks at nearby stations ($\pm2$\,s) and is thus supported by network coherence rather than single-trace amplitude (Supplementary Fig.~\ref{fig:sfig_picks}a). Among the unassociated picks, both models have a similar, small number that are adjacent to a located event or coincident across stations, but the isolated remainder runs at 425 per hour for EQNet2 against 1,174 for PhaseNet+ and varies little between day and night (Supplementary Fig.~\ref{fig:sfig_picks}b), marking it as a steady detector background unrelated to the sequence. Raising the pick-score threshold clears some of this background but barely changes the gap in association ratio. EQNet2's located picks therefore rest on coherent waveform evidence, whereas most of PhaseNet+'s larger raw output consists of isolated detections that never form events.

\FloatBarrier

\suppnote{Additional visualizations of learned RPBs and attention maps}{sec:rpb_supp}

\begin{figure}[p]
    \centering
    \includegraphics[width=0.91\linewidth]{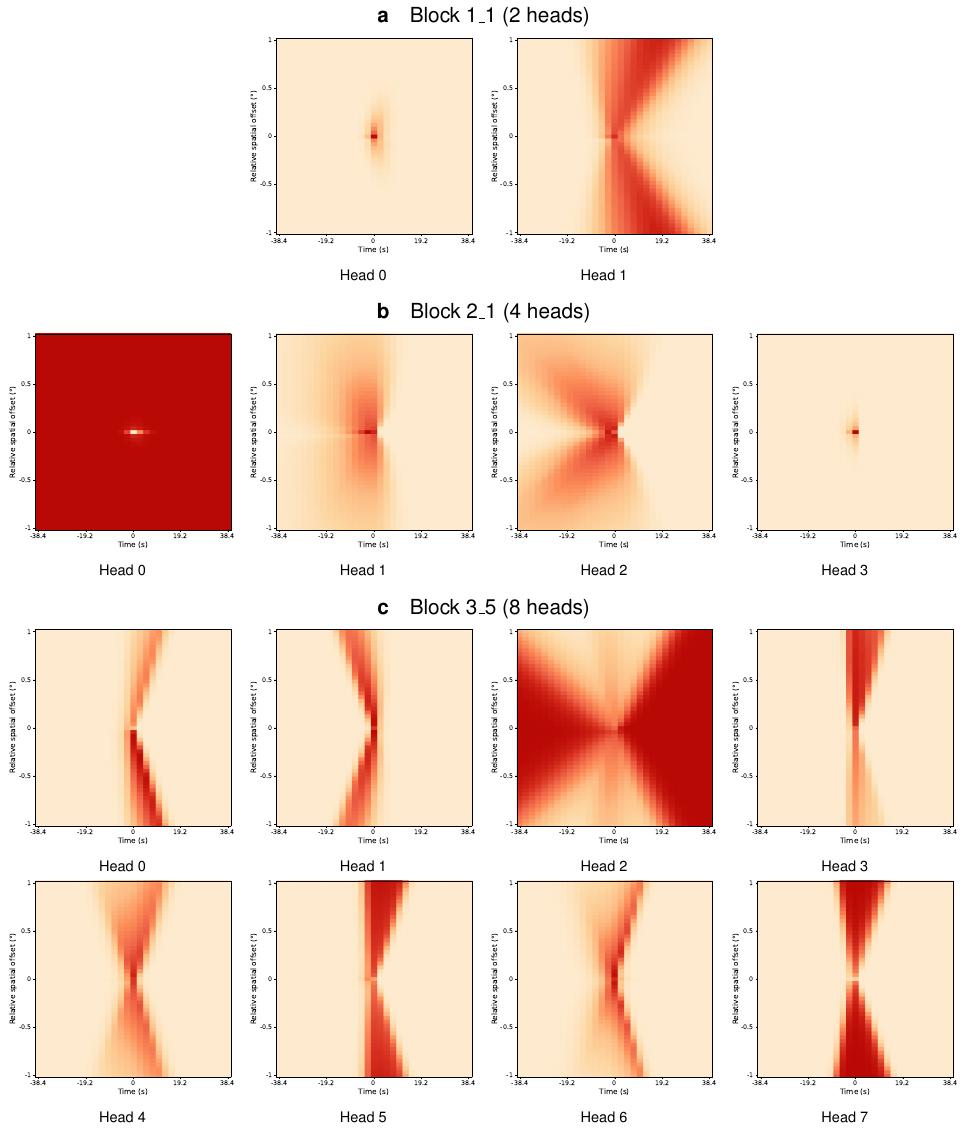}
    \caption{Relative position bias for all attention heads in the shallow and intermediate backbone stages. Panels \textbf{a--c} show blocks 1\_1, 2\_1, and 3\_5, respectively. The horizontal axes denote relative time, the vertical axes denote relative spatial offset, and darker colors indicate larger bias within each head. Because the bias ranges differ among heads, with approximate ranges of 0--8 or 0--16, every panel is normalized independently.}
    \label{fig:rpb_heads_shallow_mid}
\end{figure}

\begin{figure}[p]
    \centering
    \includegraphics[width=\linewidth]{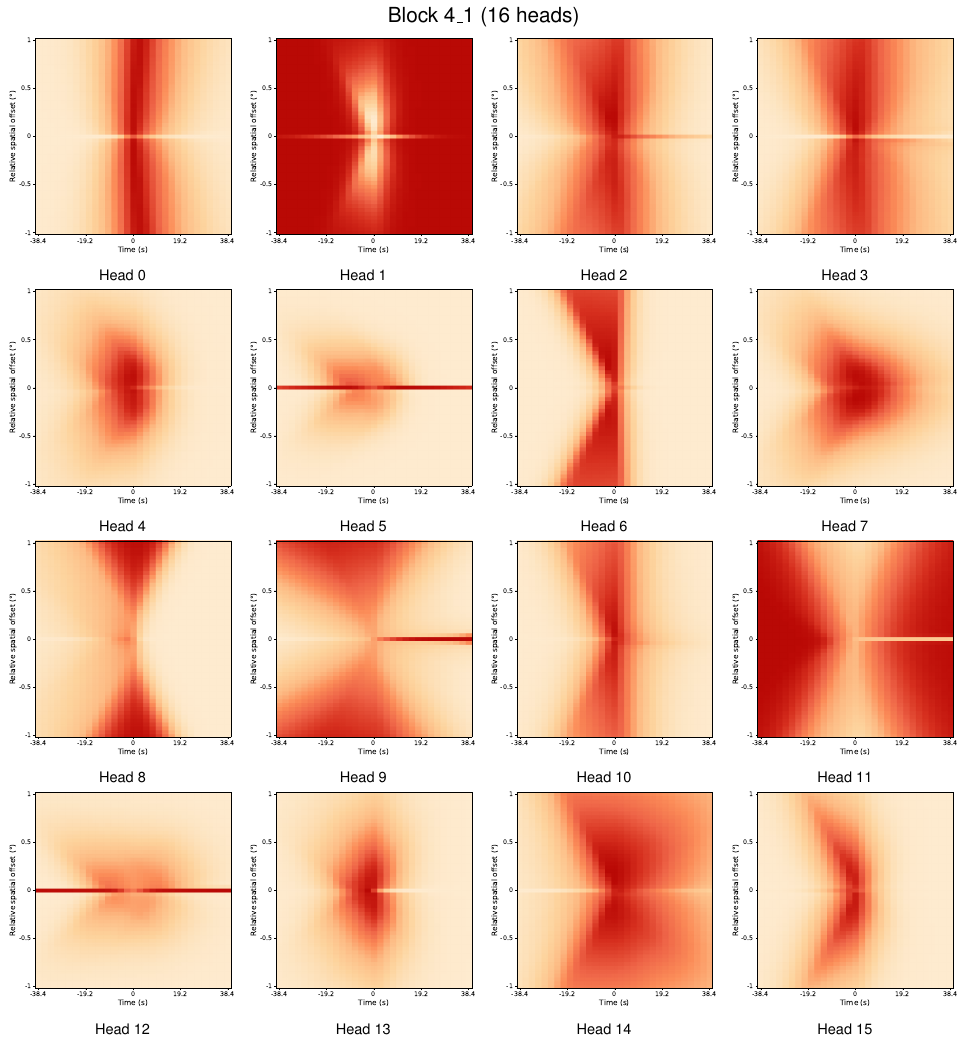}
    \caption{Relative position bias for all 16 attention heads in the deepest backbone stage (block 4\_1). The horizontal axes denote relative time, the vertical axes denote relative spatial offset, and darker colors indicate larger bias within each head. The diversity of patterns suggests specialization among attention heads. Because the bias ranges differ among heads, with approximate ranges of 0--8 or 0--16, every panel is normalized independently.}
    \label{fig:rpb_heads_deep}
\end{figure}

To examine how cross-station relationships are represented in the backbone, we visualize the learned relative position bias (RPB) for every attention head in the four backbone stages (Supplementary Figs.~\ref{fig:rpb_heads_shallow_mid} and~\ref{fig:rpb_heads_deep}). The heads exhibit distinct patterns, including spatially localized, temporally broad, and slanted structures. Some of the slanted structures are consistent with moveout-like time--distance relationships, but the pattern is not shared by all heads. The bias ranges fall into two approximate groups, 0--8 and 0--16. Each head is therefore normalized independently for visualization.

\begin{figure}[p]
    \centering
    \begin{subfigure}{\linewidth}
        \centering
        \includegraphics[width=0.82\linewidth]{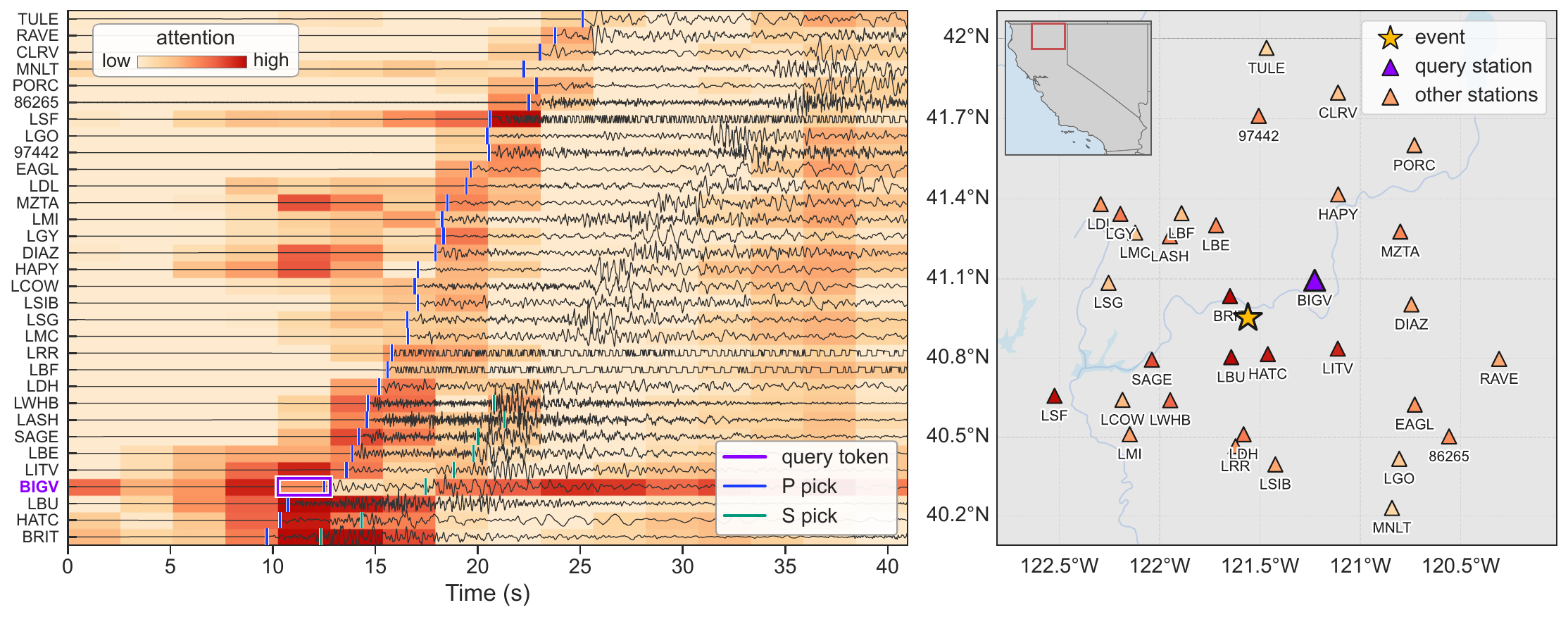}
        \caption{Query station BK.BIGV.00.HH.}
    \end{subfigure}

    \begin{subfigure}{\linewidth}
        \centering
        \includegraphics[width=0.82\linewidth]{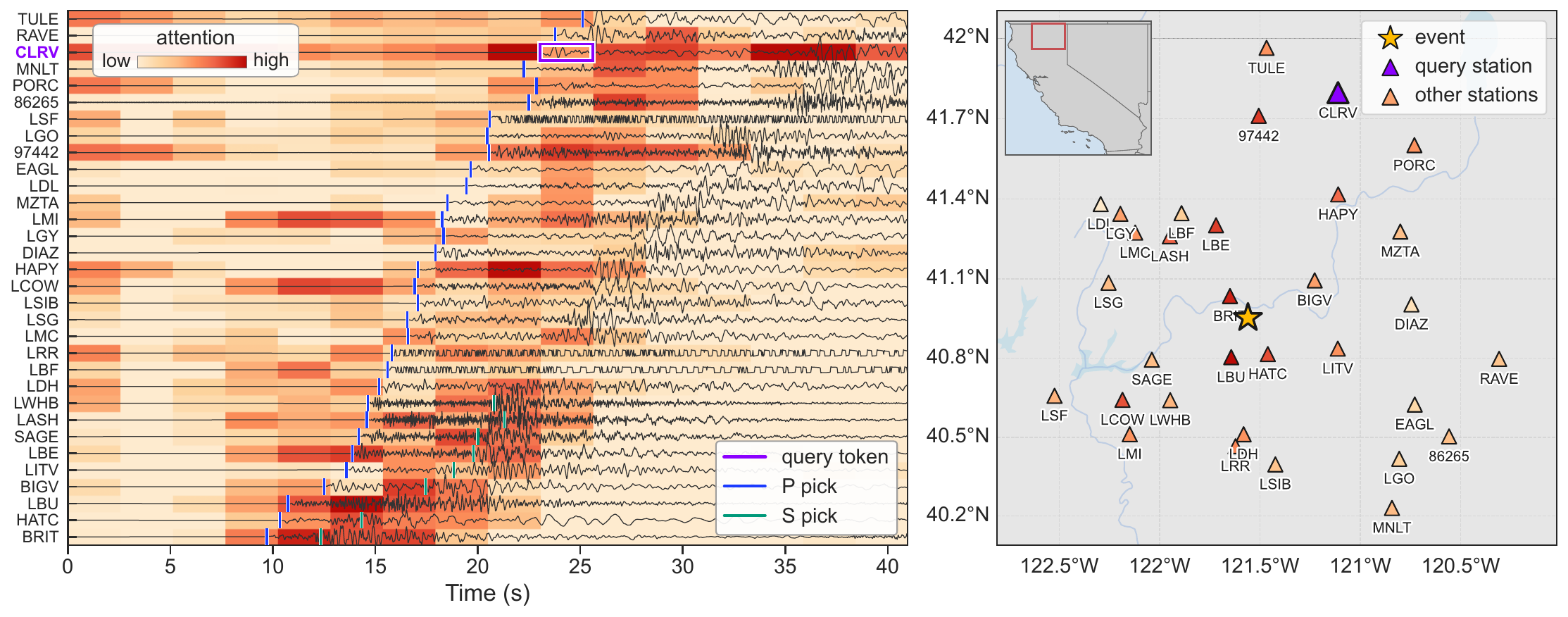}
        \caption{Query station BK.CLRV.00.HH.}
    \end{subfigure}

    \begin{subfigure}{\linewidth}
        \centering
        \includegraphics[width=0.82\linewidth]{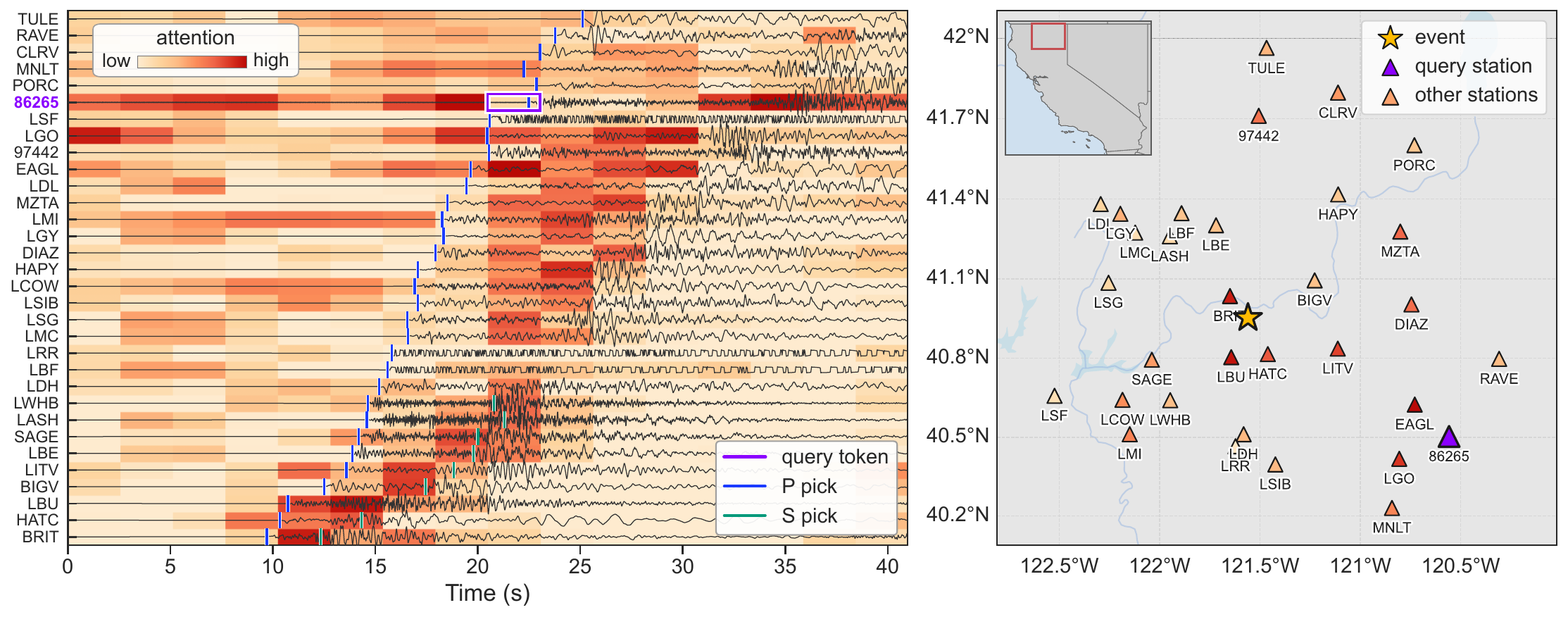}
        \caption{Query station CE.86265.10.HN.}
    \end{subfigure}
    \caption{Additional attention maps for event nc73934641 from the deepest Swin Transformer block. The three query stations were selected to represent different positions within and near the edge of the station network. The token containing the query P arrival is enclosed by a purple rectangle, the query station is indicated by a purple triangle, and darker colors indicate higher attention weights.}
    \label{fig:attention_maps_additional}
\end{figure}

We additionally compare attention maps from the deepest backbone stage for three query stations located in different parts of the network (Supplementary Fig.~\ref{fig:attention_maps_additional}). All three maps are from the same event and are intended to illustrate how attention varies with query-station geometry rather than to evaluate generalization across events.

\FloatBarrier

\suppnote{Computational cost evaluation}{sec:compute_cost}

GFLOPs of models are calculated for inputs comprising 16 stations, three components, and 4096 time samples. The time and memory costs are tested using an NVIDIA RTX 6000 Ada GPU with data from the NCEDC 2023 dataset. The sublinear scaling of runtime with the number of traces for PhaseNet+ and EQNet2 may reflect code execution overhead. Association ratios are calculated for the Berkeley earthquake case with different EQNet2 model sizes and station-window sizes for attention. The association ratio is defined as the number of picks after association and location divided by the number of raw picks.

\begin{table}[htbp]
\centering
\caption{Computational cost and performance scaling statistics. EQNet2-Base corresponds to the default EQNet2 model used in the main-text experiments.}
\label{tab:compute_cost}

\small
\setlength{\tabcolsep}{5pt}

\begin{tabular}{
llccccc
}
\toprule
Category & Metric
& {PhaseNet+}
& {EQNet2-Tiny}
& {EQNet2-Small}
& {EQNet2-Base}
& {PhaseNO} \\
\midrule

Model size
& {GFLOPs}
& 2.72 & 3.49 & 10.69 & 37.75 & 167.23 \\

& {Params (M)}
& 1.55 & 1.02 & 4.02 & 15.97 & 2.37 \\

\midrule

Runtime
& {Time (ms), 16 sta.}
& 3.7 & 24.4 & 24.9 & 31.9 & 167.2 \\

& {VRAM (GB), 16 sta.}
& 0.14 & 0.25 & 0.27 & 0.42 & 6.88 \\

& {Time (ms), 32 sta.}
& 5.1 & 26.9 & 29.3 & 54.9 & 685.3 \\

& {VRAM (GB), 32 sta.}
& 0.27 & 0.49 & 0.53 & 0.76 & 27.25 \\

\midrule

P picking
& {Precision}
& 0.988 & 0.981 & 0.982 & 0.982 & 0.942 \\

& {Recall}
& 0.993 & 0.996 & 0.996 & 0.997 & 0.990 \\

& {F1}
& 0.991 & 0.988 & 0.989 & 0.989 & 0.965 \\

\midrule

S picking
& {Precision}
& 0.967 & 0.957 & 0.960 & 0.961 & 0.937 \\

& {Recall}
& 0.959 & 0.976 & 0.979 & 0.981 & 0.972 \\

& {F1}
& 0.963 & 0.966 & 0.970 & 0.971 & 0.954 \\

\bottomrule
\end{tabular}
\end{table}

\begin{table}[t]
    \centering
    \caption{
    Catalog-level association ratio (\%) in the Berkeley earthquake case for different EQNet2 model sizes and station-window sizes for attention.
    Each cell reports the association ratio, followed by the number of located picks
    over the number of raw picks (located/raw). The 69-station configuration uses
    the full station set.
    }
    \label{tab:association_ratio_scaling}
    \renewcommand{\arraystretch}{1.20}
    \setlength{\tabcolsep}{10pt}
    \begin{tabular}{cccc}
        \toprule
        \makecell[c]{Station\\window size}
        & EQNet2-Tiny
        & EQNet2-Small
        & EQNet2-Base \\
        \midrule

        18
        & \makecell[c]{6.06\%\\{\scriptsize 2,801 / 46,206}}
        & \makecell[c]{7.68\%\\{\scriptsize 2,967 / 38,625}}
        & \makecell[c]{9.33\%\\{\scriptsize 2,683 / 28,762}} \\

        36
        & \makecell[c]{8.66\%\\{\scriptsize 2,976 / 34,361}}
        & \makecell[c]{11.24\%\\{\scriptsize 3,139 / 27,921}}
        & \makecell[c]{12.73\%\\{\scriptsize 3,020 / 23,727}} \\

        69
        & \makecell[c]{13.06\%\\{\scriptsize 3,417 / 26,171}}
        & \makecell[c]{17.60\%\\{\scriptsize 3,838 / 21,803}}
        & \makecell[c]{22.06\%\\{\scriptsize 3,501 / 15,871}} \\

        \bottomrule
    \end{tabular}
\end{table}

\FloatBarrier
\clearpage
 \end{document}